\documentclass[usenatbib]{mn2e}
\usepackage{color,hyperref}
\definecolor{darkblue}{rgb}{0.0,0.0,0.3}
\hypersetup{colorlinks,breaklinks,
            linkcolor=darkblue,urlcolor=darkblue,
            anchorcolor=darkblue,citecolor=darkblue}
\usepackage{amsmath,amssymb}

\usepackage{txfonts}
\DeclareSymbolFont{cmletters}{OML}{cmm}{m}{it}
\DeclareMathSymbol{v}{\mathalpha}{cmletters}{"76}

\usepackage{graphicx}

\defcitealias{mtb12b}{MTB12a}
\newcommand{\acknowledgements}{\section*{Acknowledgements}}
\newcommand{\hr}{h_{0.3}} 
\newcommand{\ubs}{Sw J1644+57}
\renewcommand{\sun}{\odot}

\newcommand{\be}{\begin{equation}}
\newcommand{\ee}{\end{equation}}

\newcommand{\bh}{{\rm \bullet}}
\newcommand{\st}{\star}
\newcommand\simless\lesssim   
\newcommand\simmore\gtrsim   
\newcommand\apj{\rmfamily{ApJ}}%
\newcommand\apjl{\rmfamily{ApJ}}%
\newcommand\aap{\rmfamily{A\&A}}%
\newcommand\mnras{\rmfamily{MNRAS}}%
\newcommand\prd{\rmfamily{Phys.~Rev.~D}}%
\newcommand\pasp{\rmfamily{PASP}}%
\newcommand\pasj{\rmfamily{PASJ}}%
\newcommand\nat{\rmfamily{Nature}}%
\begin{document}
\label{firstpage}

\title[Sw J1644+57 gone MAD: case for dynamically-important BH magnetic flux]{Swift J1644+57 gone MAD: the case for dynamically-important magnetic flux threading the black hole in a jetted tidal disruption event}
\author[A.~Tchekhovskoy, B.~D.~Metzger, D.~Giannios, and L.~Z.~Kelley]{Alexander Tchekhovskoy$^1$\thanks{\hbox{E-mail:
      atchekho@princeton.edu~(AT)}},
Brian D. Metzger$^2$,
Dimitrios Giannios$^3$, and
Luke Z. Kelley$^4$
\\
$^1$Center for Theoretical Science, Jadwin Hall, Princeton University,
Princeton, NJ 08544, USA; Center for Theoretical Science Fellow\\
$^2$Department of Physics, Columbia University, 538 West 120th Street,
704 Pupin Hall, New York, NY 10027, USA\\
$^3$Department of Physics, Purdue University, 525 Northwestern Avenue,
West Lafayette, IN 47907, USA\\
$^4$Astronomy Department, Harvard University, 60 Garden Street MS 10,
Cambridge, MA 02138, USA}

\date{Accepted . Received ; in original form }
\pagerange{\pageref{firstpage}--\pageref{lastpage}} \pubyear{2012}
\maketitle

\begin{abstract}
  The unusual transient Swift J1644+57 likely resulted from a collimated
  relativistic jet, powered by the sudden onset of accretion onto a
  massive black hole (BH) following the tidal disruption (TD) of a
  star.  However, several mysteries cloud the interpretation of this
  event, including (1) the extreme flaring and `plateau' shape of the
  X-ray/$\gamma$-ray light curve during the first $t - t_{\rm trig}
  \sim 10$ days after the $\gamma-$ray trigger; (2) unexpected
  rebrightening of the forward shock radio emission at $t - t_{\rm
    trig} \sim$ months; (3) lack of obvious evidence for jet
  precession, despite the misalignment typically expected between the
  angular momentum of the accretion disk and BH; (4) recent abrupt
  shut-off in the jet X-ray emission at $t - t_{\rm trig} \sim 1.5$
  years.  Here we show that all of these seemingly disparate mysteries
  are naturally resolved by one assumption: the presence of strong
  magnetic flux $\Phi_{\bh}$ threading the BH.  Just after the TD
  event, $\Phi_{\bh}$ is dynamically weak relative to the high rate of
  fall-back accretion $\dot{M}$, such that the accretion disk (jet)
  freely precesses about the BH axis = our line of site.  As $\dot{M}$
  decreases, however, $\Phi_{\bh}$ becomes dynamically important,
  leading to a state of `magnetically-arrested' accretion (MAD).  MAD
  naturally aligns the jet with the BH spin, but only after an
  extended phase of violent rearrangement (jet wobbling), which in
  Swift J1644+57 starts a few days before the $\gamma$-ray trigger and explains the erratic
  early light curve.  Indeed, the {\it entire} X-ray light curve can
  be fit to the predicted power-law decay $\dot{M} \propto
  t^{-\alpha}$ ($\alpha \simeq 5/3-2.2$) if the TD occurred a few weeks
  prior to the $\gamma$-ray trigger.  Jet energy directed away from
  the line of site, either prior to the trigger or during the jet
  alignment process, eventually manifests as the observed radio
  rebrightening, similar to an off-axis (orphan) gamma-ray burst
  afterglow.  As suggested recently, the late X-ray shut-off occurs
  when the disk transitions to a geometrically-thin (jet-less) state
  once $\dot{M}$ drops below $\sim$the Eddington rate.  We predict
  that, in several years, a transition to a low/hard state will mark a revival of the jet and its associated X-ray emission. We use our model for Swift J1644+57 to constrain the
  properties of the BH and disrupted star, finding that a solar-mass
  main sequence star disrupted by a relatively low mass $M_{\bh} \sim
  10^{5}-10^{6}M_{\odot}$ BH is consistent with the data, while a WD
  disruption (though still possible) is disfavored.  The magnetic flux
  required to power Swift J1644+57 is much too large to be supplied by the star
  itself, but it could be collected from a quiescent `fossil'
  accretion disk that was present in the galactic nucleus prior to the
  TD.  The presence (lack of) of such a fossil disk could be a deciding
  factor in what TD events are accompanied by powerful jets.
\end{abstract}

\begin{keywords}
MHD -- black hole physics -- gamma-rays: galaxies -- X-rays: galaxies -- accretion, accretion discs
\end{keywords}

\section{Introduction}

\begin{figure}
\begin{center}
    \includegraphics[width=1\columnwidth]{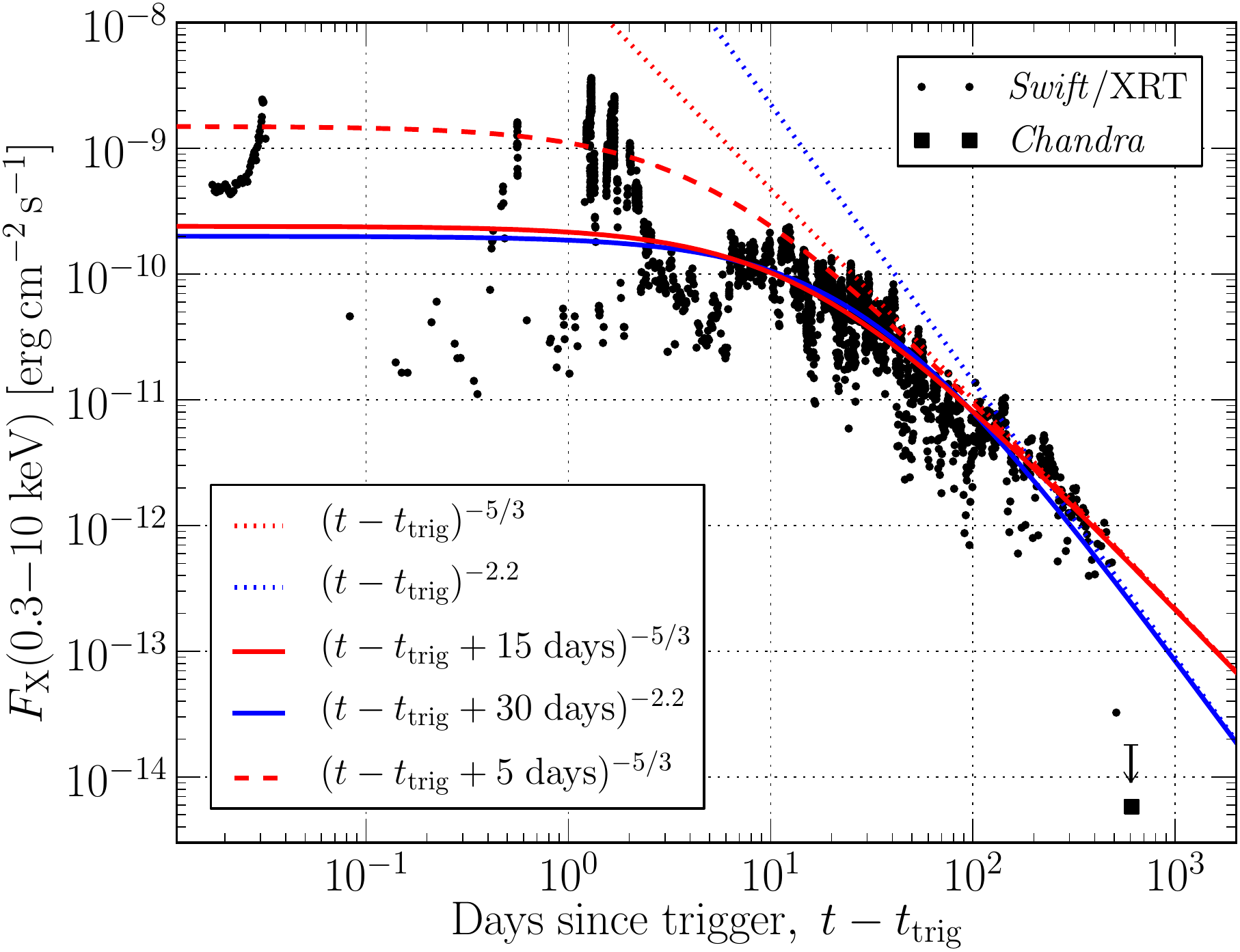}
  \end{center}
  \caption{X-ray light curve of \ubs, as measured by the {\it Swift} X-ray Telescope (XRT) and {\it Chandra}.  When plotted as a function of time since the $\gamma-$ray trigger $t - t_{\rm trig}$ the average emission shows a `plateau-like' phase lasting for $\sim 10$~days, which naively appears inconsistent with the predicted power-law decay $\dot{M}_{\rm fb} \propto (t - t_{\rm disr})^{-\alpha}$ in the rate of fall-back accretion following a tidal disruption event ($t = t_{\rm disr}$).  However, if the trigger time is delayed with respect to the disruption, then a `plateau'-like shape in $F_{x}(t-t_{\rm trig})$ is naturally produced ($\S\ref{sec:early-time-plateau}$). The two solid curves show $\dot M_{\rm fb}\propto t^{-\alpha}$ (arbitrary normalization) for complete ({\it red}, $\alpha = 5/3$ for trigger time delay $t_{\rm trig}-t_{\rm disr} = 15$ days)  and partial stellar disruption ({\it blue}, $\alpha=2.2$, $t_{\rm trig}-t_{\rm disr} = 30$ days), while the dotted curves show the conventional versions of power-law fits that neglect the trigger time delay.  The trigger time delay for the solid lines are chosen to match $\dot{M}_{\rm fb}$ with the the average luminosity of the early plateau phase ($t - t_{\rm trig} \lesssim 10$ days).   If we instead match $\dot{M}_{\rm fb}$ to the `envelope' created in the light curve by the brightest flares, then adopting a shorter trigger time delay is also consistent with the data, e.g., $t_{\rm trig}-t_{\rm disr}=5$ days for a complete disruption ({\it red dashed line}, $\alpha=5/3$).  This timescale is consistent with the first evidence for activity from \ubs~(\citealt{Burrows+11,Krimm+11,Zauderer+11}) 4 days prior to the first BAT trigger.}
  \label{fig:Fx}
\end{figure}

The unusual soft $\gamma$-ray/X-ray/radio transient Swift
J164449.3+573451 (hereafter \ubs) has been broadly interpreted as
resulting from a relativistic outflow, powered by accretion following
the tidal disruption (TD) of a star by a massive black hole (BH) (\citealt{Bloom+11,Burrows+11,Levan+11,Zauderer+11}).  
Evidence supporting this model includes the rapid onset of \ubs\ and its
location at the center of a compact galaxy at redshift $z
\simeq 0.353$ (\citealt{Levan+11}).  At least until recently, the SED showed two
distinct components, which led \citet{Bloom+11} and \citet{Burrows+11} to suggest different sources for the X-ray and radio
emission (see also \citealt{Liu+12}).  The X-ray/$\gamma$-ray
emission is highly variable, which indicates an origin from relatively
small radii, likely from a location internal to the jet itself (although see \citealt{Socrates12}).
Figure~\ref{fig:Fx} shows that the emission was particularly variable for the first $\sim10$ days after the \emph{Swift}/BAT
trigger, $t_{\rm trig}$ (though roughly constant in a time-averaged sense), after which point undergoing a power-law decline,
\begin{equation}
L_{X} \propto (t-t_{\rm trig})^{-\alpha},
\label{eq:Lx}
\end{equation}
with $\alpha \sim 5/3$, consistent with the rate of fall-back accretion in simple TD
models (e.g.~\citealt{Rees88}; \citealt{Lodato+09}; \citealt{Guillochon&Ramirez-Ruiz12}; \citealt{Stone+12}; although
see \citealt{Cannizzo+11}). The X-ray flux has recently abruptly
dropped by more than two orders of magnitude, indicating that the jet has apparently `shut off' approximately 500 days after
the initial trigger \citep{Zauderer+12}. The total isotropic X-ray
energy radiated to date is $\sim 5\times 10^{53}$ ergs.

In contrast to the X-ray emission, brightness temperature constraints
place the radio emission from \ubs\ at much larger radii, suggesting
that it instead results from synchrotron emission powered by the shock
interaction between the relativistic jet and the surrounding
circumnuclear medium \citep{Giannios&Metzger11,Zauderer+11,Metzger+12,Berger+12,Wiersema+12,Zauderer+12}. By
modeling the observed radio emission based on the first several weeks of data, \citet*{Metzger+12} (hereafter MGM12) derived
values for the bulk Lorentz Factor $\Gamma_j \approx 10$, opening
angle $\theta_{\rm j} \sim 1/\Gamma_{j} \sim 0.1$, and beaming
fraction $f_{\rm b} \approx 3\times 10^{-3}$ of the jet which are
remarkably similar to those of AGN jets.
\citet{Berger+12} (hereafter B12) presented updated radio light
curves of \ubs, which showed a distinct {\it rebrightening} starting at $t-t_{\rm trig}\sim 1$ month and peaking
on a timescale of several months. This behavior is surprising since the emission is significantly brighter than expected if the blast
wave were evolving with a relatively constant energy, as would be expected if the instantaneous jet power tracked the X-ray light curve.
B12 proposed that this large additional energy results from slower material catching up to the forward
shock at late times. Regardless of its interpretation, however, the radio rebrightening clearly indicates the jet structure
(angular or temporal) is more complex than those commonly and
successfully applied to normal gamma-ray burst afterglows \citep{pk02,Cao&Wang12,Liu+12}.

The discovery of a jetted TD event presents several
theoretical mysteries.  Relativistic jets from AGN are thought to result from magnetic, rather
than hydrodynamic, collimation and acceleration (e.g.~\citealt{Rees+82}).\footnote{One theoretical objection to a hydrodynamic jet is the drag due to Compton cooling (e.g.~\citealt{Phinney82}), which could in principle, however, be substantially reduced
  in the case of a TDE due to the lack of a previously-existing broad line region.} If the jet energy derives from the Penrose-Blandford-Znajek process, then the total jet power is given by \citep{tch10a}:
\begin{eqnarray}
P_{j} 
&=& \frac{\kappa c}{16\pi r_g^2} \Phi_{\bh}^{2} \omega_{\rm H}^2
f(\omega_{\rm H}) \nonumber \\
\label{eq:pjPhi30}
&=& 1.2\times 10^{47}\Phi_{\bh,30}^{2} M_{\bh,5}^{-2}\, \omega_{\rm H}^2 f(\omega_{\rm H})\ {\rm erg\,s^{-1}},\\ 
&=& 0.5\times 10^{47}\Phi_{\bh,30}^{2} M_{\bh,5}^{-2}\ {\rm erg\,s^{-1}}, 
\label{eq:pj}
\end{eqnarray}
where $\kappa\approx0.045$, $r_g=GM_\bh/c^2$ is BH gravitational radius;
$\Phi_{\bh} = 10^{30}\Phi_{\bh,30}$ cgs is the magnetic flux threading the hole;
$M_{\bh} = M_{\bh,5}10^{5}M_{\sun}$ is the BH mass; $\omega_{\rm
  H}=a/[1+(1-a^2)^{1/2}]$ is dimensionless angular frequency of BH
horizon (equals unity for a maximally spinning BH); and $f(\omega_{\rm
  H})=1+0.35\omega_{\rm H}^2-0.58\omega_{\rm H}^4$ is a high-spin
correction, while the normalization in the third line has been
calculated for $a = 0.9$.

\ubs~radiated $\sim 2\times 10^{53}$ ergs over the first
$\sim 10$ days after the trigger, corresponding to an average isotropic X-ray
luminosity $L_{\rm X}^{\rm trig} \sim 2\times 10^{47}$ erg s$^{-1}$. The total (true) jet power during this interval was thus $P_{j}^{\rm trig} = 2(f_{b}\epsilon_{\rm bol}\epsilon_{X}^{-1})L_{\rm X} \approx 10^{46}(f_{\rm
  b}\epsilon_{\rm bol}\epsilon_{X}^{-1}/0.03)$ erg s$^{-1}$, where $\epsilon_{\rm b} \sim$ few is a bolometric correction; $\epsilon_{X}< 1$ is the jet radiative efficiency; and the factor of 2 accounts for the other jet beamed away from Earth.  Equation (\ref{eq:pj}) shows that in order to explain $P_{j}^{\rm trig}$, the required magnetic flux $\Phi_{\bh,30}
\sim M_{\bh,5}$ for $a = 0.9$, is several orders of magnitude larger than that through a typical main sequence star \citep{Bloom+11}.
In $\S\ref{sec:flux}$ we discuss possible alternative sources of magnetic flux, such as could be supplied by a pre-existing quiescent accretion disk.

Regardless of its origin, the high luminosity of \ubs\ requires a large magnetic flux.  In fact, at least two independent arguments suggest that such magnetic flux was actually present.  First, note that a significant fraction of the mass of the disrupted star, and hence of
its magnetic flux or that of a quiescent disk, is accreted on the characteristic fall-back time $t_{\rm fb}$ [eq.~(\ref{eq:tfb})] of
the most tightly bound tidal debris (e.g.~\citealt{Ulmer1999}; \citealt{Strubbe&Quataert09}).  Equation (\ref{eq:pj}) would thus
naively imply that the average jet power should be constant, or rising, at times $t \gg t_{\rm fb}$, in contradiction to the observed power-law decline
in the X-ray luminosity. Using 3D GRMHD simulations, \citet{tch11} have shown that if the magnetic flux $\Phi_{\bh}$ is sufficiently high, then magnetic forces
impede accretion onto the BH, causing the flow to enter a
`magnetically-arrested' (MAD; e.g. \citealt{nia03}; \citealt{Igumenshchev08}). MAD
flows achieve a quasi-steady state as the result of 3D instabilities, which allow matter to slip past the field lines towards the BH.  This process regulates the jet power [eq.~(\ref{eq:pj})] to be proportional to the BH feeding rate,
$P_{j} \propto \dot{M}$ [eq.~(\ref{eq:Pjet})]. {\it Hence, the fact that the late-time jet power in \ubs\ faithfully tracks the expected rate of fall-back accretion $\dot{M} \propto t^{-\alpha}$ is only naturally understood if the flow is in a magnetically-arrested state}.

Additional evidence for a strong magnetic flux is related to the mystery raised by \citet{Stone&Loeb12}, who
noted that in general a TD jet should precess
if it is pointed along the angular momentum axis of the accretion disk.\footnote{A disk-aligned jet is expected if outflows from the disk are sufficiently powerful to direct the jet on large scales, as is likely if the accretion rate is highly super-Eddington.} Lack of clear evidence\footnote{\citet{Saxton+12} and \citet{Lei+12} argue that the `dipping' behavior and other quasi-periodic features observed in the X-ray light curve of \ubs~are due to the jet precession/nutation.  However, the high duty cycle of observed emission would still require a relatively small (and hence fine-tuned) inclination between the BH spin and stellar orbit ($\lesssim 10-20^{\circ}$).} for large-scale jet precession in \ubs\ thus requires either a set of highly unlikely circumstances, such as an unphysically low BH spin or near-perfect
alignment between the angular momentum of the BH and the original orbit of the disrupted star \citep{Stone&Loeb12}, or some mechanism
for aligning the angular momentum of the disk with the BH spin.  In fact, recent numerical simulations by \citet{mtb12b} (hereafter \citetalias{mtb12b}) show that such an alignment between the disk and BH spin axis {\it can} occur due to MHD forces, but only if the strength of the magnetic field threading the BH is similar to that required for MAD accretion.  If such an alignment process occurred in \ubs, then {\it the lack of observed precession also provides indirect evidence for a strong magnetic flux.}

In this paper, we present a new physical scenario for \ubs\ which addresses the seemingly disparate mysteries raised above, including the shape of the X-ray/$\gamma$-ray light curve; lack of jet precession; and late radio rebrightening.  We show that all of these features are naturally expected given a single assumption: the presence of a strong magnetic flux threading the BH. We begin in $\S\ref{sec:phenom}$ with some basic phenomenological considerations.  Then in $\S\ref{sec:scenario}$ we overview the timeline of our proposed scenario for \ubs. In $\S\ref{sec:constraints}$ we use our model to constrain the properties of the BH and stellar progenitor.   In $\S\ref{sec:discussion}$ we discuss our results, including the nature of the disrupted star ($\S\ref{sec:progenitor}$); the origin of the magnetic flux ($\S\ref{sec:flux}$); the nature and duration of the flaring state ($\S\ref{sec:flaring}$); the origin of the radio rebrightening ($\S\ref{sec:radio}$); and future predictions ($\S\ref{sec:predictions}$). We present our conclusions in $\S\ref{sec:conclusions}$. Throughout the paper we use Gaussian-cgs units and set the zero time to the point of disruption, $t_{\rm disr}=0$.

\begin{figure*}
\begin{center}
    \includegraphics[width=\textwidth,clip=True]{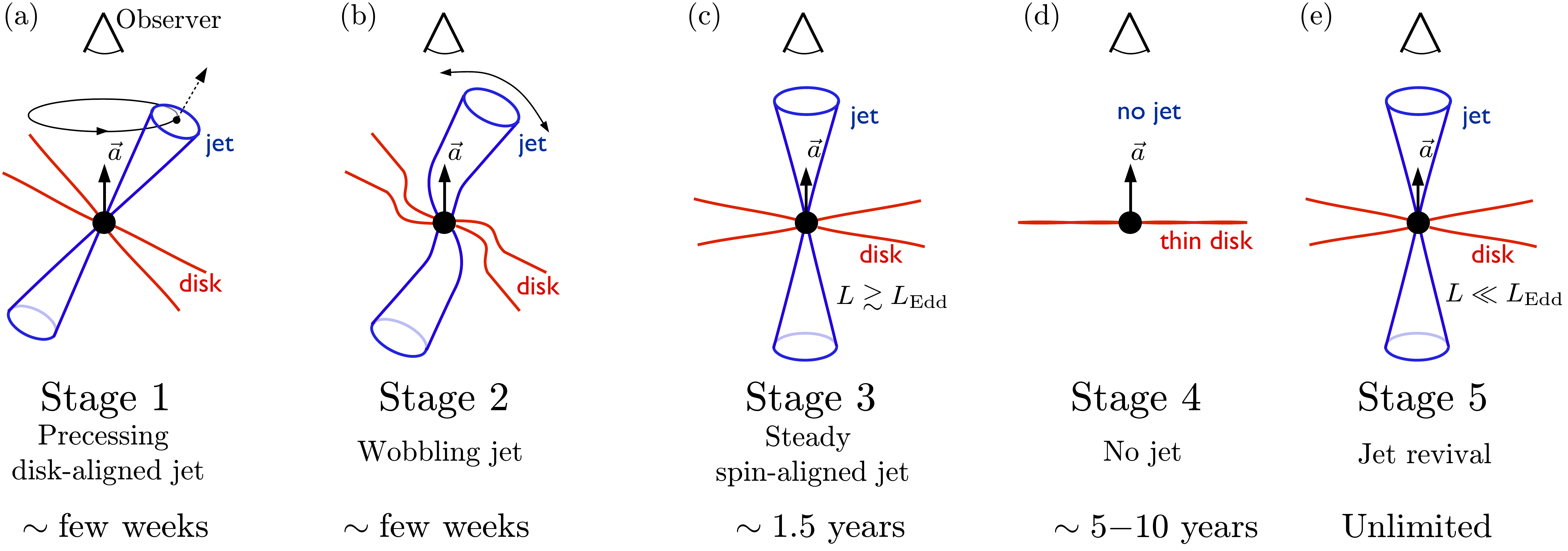}
  \end{center}
  \caption{Stages in the proposed model for \ubs. {\bf [Panel (a)]: Stage 1.} Shortly
after the stellar disruption, the magnetic flux threading the BH
$\Phi_{\bh}$ increases, as it is dragged inward by the accreting material.
Stellar debris returning to the BH forms a \emph{tilted} accretion disk,
with a rotation axis that is misaligned with the BH spin.  The latter
points towards our line of site (shown as vertical in the figure).  At
early times, when the mass accretion rate $\dot{M}$ is highest, the magnetic flux $\Phi_{\bh}$ is
not dynamically important. Under these conditions, the disk undergoes
precession due to Lense-Thirring torques \citep{fragile07} and the jets
point along the disk axis. Their high-energy emission is beamed away from
Earth and is not detectable.  {\bf [Panel~(b)]: Stage 2.} As $\dot{M}$
decreases with time, the magnetic flux eventually becomes dynamically
important, leading to a state of magnetically-arrested disk (MAD)
\citep{nia03,tch11}.  In the MAD state, the magnetic field is sufficiently
strong to offset gravitational forces acting on the inner disk and to
align the axis of the disk axis (and hence the jets) with the BH spin
(so-called ``magneto-alignment effect''; \citetalias{mtb12b}). However, it
takes time for the entire disk and jet to align with the BH spin.  As
$\dot{M}$ decreases, the BH magnetic flux and jet power $P_j \propto \dot
M$ also decrease. As excess magnetic flux leaks out of the BH into the
disk, MAD encompasses a larger and larger fraction of the disk.  During
this process, the jet pushes against the disk and wobbles erratically (see
movies in \citetalias{mtb12b}).  As the jet comes in and out of our line of
sight, this produces large amplitude variations ("flares") in lightcurve
of \ubs\ during the first $\sim10^6$ seconds after the trigger (see
Fig.~\ref{fig:Fx}). {\bf [Panel (c)]: Stage 3.}  Once MAD encompasses the
entire disk, the disk/jet alignment with the BH spin completes.  Since the
jet direction points steadily towards Earth, its X-ray emission becomes
less variable as it continues to track the rate of fall-back accretion. 
{\bf [Panel~(d)]: Stage 4.}  Once $\dot{M}$ deceases below $\sim30$\% of
the Eddington accretion rate $\dot{M}_{\rm Edd}$, the disk transitions to
a geometrically-thin state and the jet shuts off, producing an abrupt
decline in X-ray emission at $t\gtrsim$ 500 days.  {\bf [Panel~(e)]: 
Stage 5.}  At very late times, once the accretion rate drops below a few
percent of $\dot{M}_{\rm Edd}$, the disk will again enter a
geometrically-thick regime, and the jet may turn back on, again analogous
to state transitions in X-ray binaries.  For \ubs~this `jet revival' is
estimated to occur between the years $\sim$ 2016-2022
($\S\ref{sec:predictions}$).}
  \label{fig:cartoon}
\end{figure*}

\section{Phenomenological Considerations}
\label{sec:phenom}

We begin by discussing the origin of several features in the X-ray and
radio light curves of \ubs\ from a phenomenological perspective. Then in
$\S\ref{sec:scenario}$ we present a more systematic overview of
our model.

\subsection{Relativistic Jet as the Origin of X-ray Emission}

Figure~\ref{fig:Fx} shows the soft X-ray (0.3-10 keV) light curve of
\ubs.  Upper limits on the mass of the host galaxy and the observed variability timescale place a rough upper limit of $M_{\bh} \lesssim
10^{7}M_{\odot}$ on the mass of the central BH (e.g.~\citealt{Bloom+11}).  The jet luminosity was thus highly super-Eddington over at least the first several days of
activity, even after correcting for jet beaming (MGM12).  At $t\gtrsim10$~days, the time-averaged emission follows
a power-law decline with temporal index $\alpha\simeq5/3$
(eq.~[\ref{eq:Lx}]), similar to the rate of mass fall-back in standard
TD scenarios \citep{Rees88}.  This decay rate has been used as
evidence that \ubs~was in fact a TD event, but as we discuss below, it
is not clear {\it a priori} why the jet power should so faithfully
track the accretion rate.  
\subsection{Early Time Light Curve `Plateau'}
\label{sec:early-time-plateau}
The first $\sim 10$ days of \ubs~was characterized by particularly intense flaring (Fig.~\ref{fig:Fx}).  The luminosity during this period, though variable by several orders of magnitude, was approximately constant on average.  Such a `plateau' has no obvious explanation in TD scenarios, but, as we now discuss, it naturally results if the $\gamma-$ray trigger was delayed with respect to the time of disruption.

Figure \ref{fig:Fx} shows that a plateau consistent with the data is reproduced simply by shifting the zero-point of time, even for a purely power-law decay in the assumed flux.  In order to reproduce the duration of the plateau and match the predicted accretion rate to the average X-ray flux, we find that a trigger delay $t_{\rm trig} - t_{\rm disr} \sim$ weeks-month is required, depending on whether the TD event was a complete ($\alpha \simeq 5/3$) or partial disruption
($\alpha = 2.2$; we discuss this distinction in $\S\ref{sec:scenario}$).  If we instead fit the `envelope' of X-ray emission set by the brightest flares (a factor $\sim 10$ times higher flux than the plateau average), then a shorter trigger delay $t_{\rm trig}-t_{\rm disr} \sim 5$ days is also consistent with the data (justification of such a possibility is discussed in \S\ref{sec:trigger}).  There in fact was evidence for jet activity
$\sim$4 days prior to the \emph{Swift} trigger (on March 28, 2011) as
seen in both \emph{Swift}/BAT data
\citep{Krimm+11,Burrows+11}, as well as inferred by the
rise time of the radio emission (\citealt{Zauderer+11}; MGM12).
However, our fits in Figure \ref{fig:Fx} show that the TD event could
have occured much earlier than this time.

\subsection{Jet Activity Prior to Trigger and Radio Rebrightening}
\label{sec:radio_rebrightening}

What could cause such a long delay before the onset of jet emission?  One possibility is that the process of jet formation requires special conditions which only became satisfied at late times after the disruption, such as the accumulation of a critical quantity of magnetic flux ($\S\ref{sec:flaring}$). It is also possible that the jet {\it was}
active soon after disruption, but that it was initially pointed away from our line of site, as could be expected if the black hole spin and the angular
momentum of the fall-back disk were initially misaligned.  Even if the
jet eventually fully aligns towards our line of site, it might not be pointed towards us at all times during initial stages in the alignment process (see below).  

Although X-ray emission from a misaligned jet is unobservable, it still imparts kinetic energy into the ambient medium surrounding the black
hole.  Synchrotron radio emission from this off-axis blast-wave would not be visible
initially due to relativistic beaming, but could become visible once
the ejecta slows to mildly relativistic Lorentz factor $\gamma\sim
2(0.5/\theta_{\rm ma}),$ where $\theta_{\rm ma}$ is the angle between the jet and line of site, normalized to a typical value $\sim 30^{\circ}$.  A blast wave of [isotropic] energy $E^{\rm
  iso} = 10^{54}E_{54}$ erg that interacts with an external medium of density $n=10n_1$ cm$^{-3}$ (see MGM12, B12 for motivation for these characteristic values) will decelerate to $\gamma=2$ at a distance $R_{\rm dec}\simeq 2.4\times 10^{18}(E_{54}/n_1)^{1/3}$cm from the BH. The observer time corresponding to emission from this radius $t_{\rm obs}\simeq R_{\rm dec}(1-\beta\cos \theta_{\rm ma})/\beta c \simless 1$yr, for $\theta_{\rm ma}\simless 0.5$.  Delayed emission from such an off-axis jet thus provides a possible explanation for the otherwise mysterious radio rebrightening observed to peak $\sim 4$ months after the onset of \ubs\ (B12; see $\S$\ref{sec:radio} for further discussion). 

\subsection{Intense Early-Time Flaring Followed by a Steady Decline}
\label{sec:earlyflaring}

Large-amplitude flares in the early $\gamma$-ray/X-ray lightcurve of \ubs~(Fig.~\ref{fig:Fx}) could result from geometric effects, such as changes in jet orientation which cause emission to periodically beam in and out of our line of sight.  This erratic `wobbling' of the jets is a natural consequence of the alignment process.  Eventually, once the jet direction becomes stable, one of the jets continuously points towards us.  This produces relatively steady X-ray emission at late times ($\gtrsim 10$ days), which decays as a power-law in time, reflecting the rate at which stellar debris returns to the BH.

\subsection{Jet Shutoff}
\label{sec:jet-shutoff}

The X-ray emission from \ubs~recently declined abruptly at $t \sim 500$ days, after which time the XRT is able to place only upper limits on the flux (Fig.~\ref{fig:Fx}). \ubs~was subsequently detected by \emph{Chandra} at at flux nearly 2 orders of magnitude lower than that just prior to the decline (\citealt{Sbarufatti+12,Zauderer+12}); however, this residual flux is consistent with it being the high energy extension of the same forward
shock synchrotron emission observed at radio frequencies (\citealt{Zauderer+12}).  Such a jet `shut off' may occur once the accretion rate drops below a fraction of the Eddington accretion rate (\citealt{Zauderer+12}), since after this time the disk becomes geometrically-thin and enters a thermally-dominant
accretion state, which are not observed to produce powerful jets \citep{2011ApJ...739L..19R}.

\begin{figure}
\begin{center}
    \includegraphics[width=1\columnwidth]{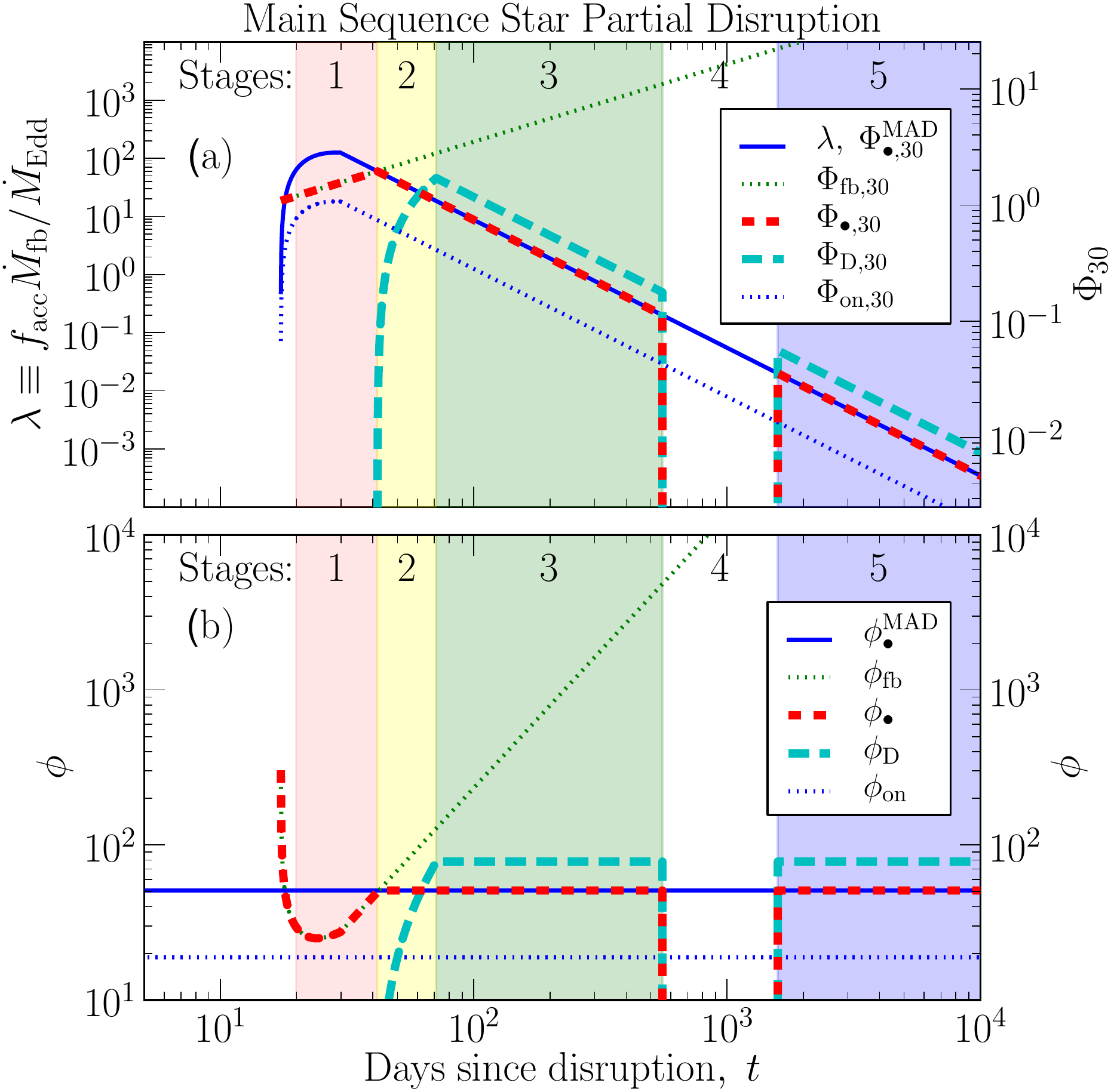}
  \end{center}
  \caption{Various quantities as a function of time since disruption,
    assuming a partial disruption of a MS star and a delay of
    $t_{\rm trig}-t_{\rm disr}=60$~days between the disruption and the
    $\gamma$-ray trigger ($\S\ref{sec:early-time-plateau}$).
    Different stages in our model for \ubs~are indicated with color
    coding and numbers ($\S\ref{sec:scenario}$; see
    Fig.~\ref{fig:cartoon} for a detailed explanation). {\bf [Panel (a)]:}
    Mass accretion rate $\lambda = \dot{M}/\dot{M}_{\rm Edd}$ as a
    fraction of the Eddington accretion rate $\dot{M}_{\rm Edd}$ ({\it
      blue solid line}; left axis), which peaks at $t \simeq 1.5t_{\rm
      fb}\simeq30$ days and subsequently decays as $\lambda\propto
    t^{-2.2}$.  The same curve (right axis) gives the maximum value of
    the magnetic flux threading the BH, $\Phi_{\bh,30}^{\rm MAD}\propto
    \lambda^{1/2}$, in units of $10^{30}$ G cm$^{2}$ ({\it red dashed
      line}). As stellar debris returns to the vicinity of the BH,
    it drags in magnetic flux from a pre-existing, ``fossil''
    accretion disk ($\S\ref{sec:flux}$).  This accumulated flux due to
    the swept-up fossil field increases with time as $\Phi_{\rm
      fb,30}\propto t^{2/3}$ ({\it green dotted
      line}; eq.~\ref{eq:fluxoft}).  At early times, most of this flux
    ends up on the BH, so $\Phi_{\bh,30}=\Phi_{\rm fb,30}$
    ({\it red dashed line}). For our choice of parameters (see
    below), BH 
    receives a substantial amount of magnetic flux early on,
    enough to
    overcome the ram pressure of the infalling gas and produce the
    jets, $\Phi_{\bh,30}>\Phi_{\rm on,30}$ ({\it red dashed
      line} lies above {\it blue dotted line}). Eventually,
    the central magnetic flux
    becomes dynamically-important, i.e.,
    $\Phi_{\bh,30}\simeq\Phi_{\bh,30}^{\rm MAD}$ ({\it red dashed
      line} crosses {\it blue solid
      line}).  Since the inner disk can only hold
    $\Phi_{\bh,30}^{\rm MAD}$ worth of flux on the hole (Stage 2),
    the rest leaks out and instead contributes to the disk flux ({\it
      cyan long-dashed line}), i.e.~$\Phi_{\rm D,30}=\Phi_{\rm
      fb,30}-\Phi_{\bh,30}^{\rm MAD}$.  However, the disk flux
    eventually also saturates (depending on its radial extent, eq.~\ref{eq:diskflux}), after
    which it also tracks the BH magnetic flux (Stage 3).  Once the
    Eddington ratio, $\lambda$, falls below a critical value
    ($\lambda_{\rm cr}=0.2$ in this figure) the accretion disk becomes
    geometrically-thin; the central BH and disk lose their magnetic
    flux; and the jets shut off (Stage 4). At a much later time, when
    $\lambda \lesssim 0.02$, the accretion disk becomes
    geometrically-thick again and produces powerful jets (Stage
    5). {\bf [Panel (b)]:} Dimensionless values (eq.~\ref{eq:phi}) of the magnetic
    flux shown in panel (a).  In this figure we have assumed
    following parameters: BH spin $a = 0.7$; BH mass $M_\bh =
    3\times10^5M_\sun$; stellar mass $M_\st = 0.44M_\sun$; fraction of
    star accreted by BH $f_{0.4}f_{\rm acc}=0.75$; Eddington fraction
    of the fossil disk $\lambda_{\rm fossil}=1.8\times10^{-3}$; and
    $P_j/L_X = f_{\rm b}\epsilon_{\rm bol}\epsilon_{X}^{-1}=
    0.05$. For computing flux accumulation timescale at early time 
    (eq.~\ref{eq:phimadfb}), we assume disk
    thickness $h/r=1$ as expected for a highly super-Eddington
    flow. Since at later times $h/r$ decreases, 
    for computing jet power we take a single representative value,
    $h/r=0.3$. \vspace{1cm}}
  \label{fig:fluxmsp}
\end{figure}
 
\begin{figure}
\begin{center}
    \includegraphics[width=1\columnwidth]{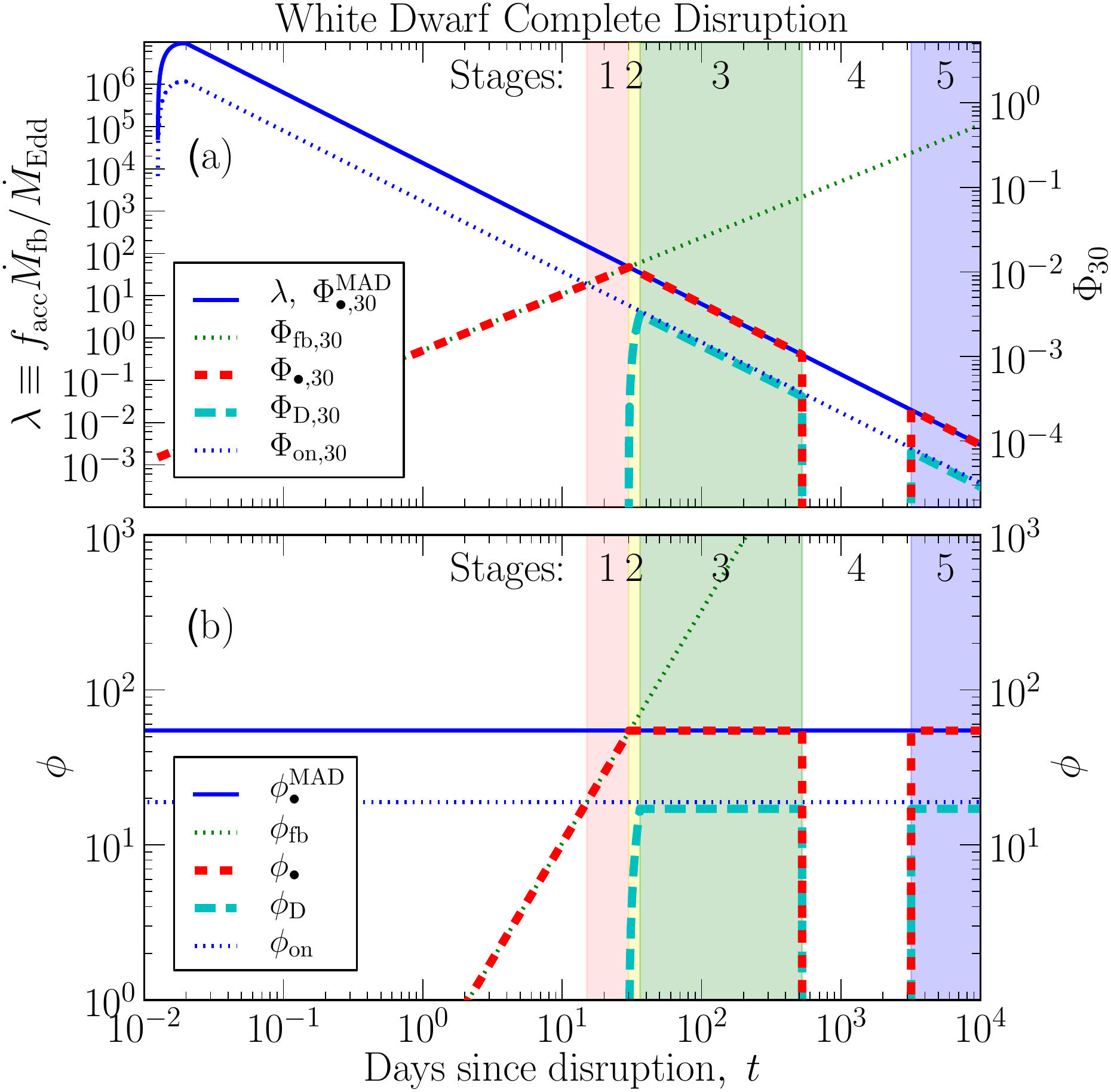}
  \end{center}
  \caption{Similar to Figure \ref{fig:fluxmsp}, but shown for the case
    of a complete disruption of a WD, assuming a disruption-trigger
    delay of $t_{\rm trig}-t_{\rm disr}=30$ days.  Note that unlike
    the MS star scenario (Fig.~\ref{fig:fluxmsp}), the mass accretion
    rate is highly super-Eddington near the peak ($\lambda \gg 100$),
    and initially jet production is suppressed (see discussion in
    $\S\ref{sec:stage-1:-precessing}$).  However, as the mass
    accretion rate decreases, $\lambda\propto t^{-5/3}$, eventually
    the magnetic flux becomes sufficiently large to overcome the ram
    pressure of the accretion flow, $\phi_\bh>\phi_{\rm on}\simeq20$ ({\it red dashed line} overtakes {\it blue dotted line}), and the jets emerge.  After this point, the jet evolution is similar to that described in Figure~\ref{fig:fluxmsp}.  In this figure, we have assumed the following parameters: $a = 0.87$; $M=10^4M_\sun$; $M_\st=0.6M_\sun$; $f_{\rm acc}=0.54$; $\lambda_{\rm fossil}=10^{-8}$; $\lambda_{\rm cr}=0.4$; and $f_{\rm b}\epsilon_{\rm bol}\epsilon_{X}^{-1} =0.003$.}
  \label{fig:fluxwdc}
\end{figure}

\section{Timeline of Proposed Scenario}
\label{sec:scenario}

We now describe the timeline for our proposed scenario for \ubs, the
stages of which are illustrated in Figure \ref{fig:cartoon}.

\subsection{Stage 0: Tidal Disruption and Flux Accumulation}

A star of mass $M_{\star}=m_\st M_\sun$ and radius $R_{\star}=r_\st R_\sun$ is tidally disrupted
by a BH of mass $M_{\bh}=10^5M_{\bh,5}M_\sun$ if its pericenter radius $R_{\rm p}$ lies
within the tidal radius $R_{\rm t} \simeq R_{\star}(M_{\bh}/M_{\star})^{1/3}$.  After disruption, approximately half of the
star is immediately unbound, while the other half remains marginally
bound and is placed on highly elliptic orbits, with the most tightly
bound material falling back on a characteristic timescale (\citealt{Stone+12})
\be
t_{\rm fb} \simeq 17.3\;{\rm d}\;M_{\bh,5}^{1/2}m_\st^{-1}r_\st^{3/2}.
\label{eq:tfallback}
\ee
The fall-back accretion rate $\dot{M}_{\rm fb}$ peaks (at  $t_{\rm
  peak}\approx 1.5t_{\rm fb}$; \citealt{Ulmer1999}) at a characteristic value
\be
\dot{M}_{\rm fb,peak} \simeq \frac{\alpha-1}{2/3}\frac{M_{\star}}{3t_{\rm fb}} 
\approx 2\times 10^{26}{\rm g\,s^{-1}}\frac{\alpha-1}{2/3}M_{\bh,5}^{-1/2}m_\st^{2}r_\st^{-3/2},
\ee
or in terms of the Eddington accretion rate $\dot{M}_{\rm Edd} \equiv L_{\rm Edd}/0.1c^{2}$,
\be
\frac{\dot{M}_{\rm fb,peak} }{\dot{M}_{\rm Edd}}  \simeq 4\times 10^{3}\frac{\alpha-1}{2/3}M_{\bh,5}^{-3/2}m_\st^{2}r_\st^{-3/2},
\label{eq:mdotpeak}
\ee and subsequently decays as a power law 
\begin{equation}
\dot{M}_{\rm fb} \simeq \dot{M}_{\rm fb,peak}\left(\frac{t}{t_{\rm fb}}\right)^{-\alpha}
\label{eq:mfb}
\end{equation}
 where $\alpha = 5/3$ for a complete disruption
and $\alpha = 2.2$ for a partial disruption (\citealt{Guillochon&Ramirez-Ruiz12}; Figs.~\ref{fig:fluxmsp}, \ref{fig:fluxwdc}).
Until recently (e.g.~\citealt{Ulmer1999,Strubbe&Quataert09}), TD models predicted that $t_{\rm fb}$ and $\dot{M}_{\rm fb,peak}$ should
depend also on the impact parameter $\beta \equiv R_{\rm t}/R_{p}$ in
addition to the stellar and BH properties; however, recent numerical
\citep{Guillochon&Ramirez-Ruiz12} and analytic \citep{Stone+12} work has
shown that these estimates were made on the faulty assumption that the
energy distribution of the disrupted stellar material is frozen-in at
pericenter instead of the tidal radius.

Matter falls back and circularizes to form an accretion disk at the radius 
\begin{equation}
R_{\rm circ} \simeq 2R_{\rm p} = 2R_{\rm t}\beta^{-1} \simeq 430  r_\st m_{\star}^{-1/3}M_{\bh,5}^{-2/3}\beta^{-1}r_{g},  \label{eq:rd}
\end{equation}
accreting soon thereafter. When the accretion rate is super-Eddington at early times, the disk may be prone to outflows driven by radiation pressure (e.g.~\citealt{Ohsuga+05,Strubbe&Quataert09}), in which case the accretion rate reaching the BH is less than $\dot{M}_{\rm fb}$. 

If the magnetic flux responsible for powering the jet in
\ubs~originates from the star itself, then a substantial fraction of
the total flux is accumulated on the relatively short fall-back time
$t \lesssim t_{\rm fb}$ (eq.~[\ref{eq:tfallback}]). However, flux
accumulation can last significantly longer if the field is swept up
from a quiescent fossil disk by the infalling debris (in which case
$\Phi_{\bh} \propto (t/t_{\rm fb})^{2/3}$; $\S\ref{sec:flux}$). On
timescales $t \gtrsim t_{\rm fb}$ the jet power $P_j \propto
\Phi_{\bh}^{2}$ [eq.~(\ref{eq:pj})] thus either saturates to a
constant value [stellar flux], or increases as $P_j \propto (t/t_{\rm
  fb})^{4/3}$ [fossil disk]. In what follows, we express the flux in a
dimensionless form, 
\begin{align}
  \label{eq:phi}
  \phi_{\bh} &\equiv \frac{\Phi_{\bh}}{(\dot M r_g^2c)^{1/2}} 
\approx 30 \Phi_{\bh,30}\left(\frac{\dot{M}}{\dot{M}_{\rm fb,peak}}\right)^{-1/2}M_{\bh,5}^{-3/4}m_\st^{-1}r_\st^{3/4},
\end{align}
that quantifies its dynamical importance in relation to the accreting gas, where $r_{g} \equiv GM_{\bh}/c^2$.  Figures \ref{fig:fluxmsp} and \ref{fig:fluxwdc} show the time evolution of the $\Phi_{\bh}$ and $\phi_{\bh}$ in our model for \ubs, based on two scenarios for the disrupted star ($\S\ref{sec:star}$).

\subsection{Stage 1: Precessing Disk-Aligned Jet (Fig.~\ref{fig:cartoon}a).}
\label{sec:stage-1:-precessing}

In general the angular momentum of the initial stellar orbit and
fall-back accretion disk will not be aligned with the spin of the BH.
Such tilted, geometrically-thick accretion disks undergo precession
due to Lense-Thirring torques, and their jets are also expected to precess
(Fig.~\ref{fig:cartoon}a). Though expected, evidence for precession is not obviously present in the light curve of \ubs~(\citealt{Stone&Loeb12}; although
see \citealt{Saxton+12, Lei+12}).  This
mystery is resolved if we postulate that our line of site is aligned with the BH spin axis ($\S\ref{sec:phenom}$), such that the emission from the tilted jet
is beamed away from us and hence is not initially detectable. 

Because the magnetic field is dynamically weakest relative to the
accretion flow ($\phi_{\bh}$ is smallest) when the accretion rate is
highest $\dot{M} \sim \dot{M}_{\rm fb, peak}$ (eq.~[\ref{eq:phi}]),
the magnetic flux does not appreciably influence the disk inclination
at these early times. This allows for a phase of jet precession as
described above. 

At even earlier times, however, when $\phi_\bh \lesssim
\phi_{\rm on}\approx20$, the ram pressure
of the quasi-spherical accretion flow (as expected for
super-Eddington accretion)
is so high that it can stifle jet
formation altogether \citep{kb09}. Since $\phi_{\bh} \propto t^{\alpha/2}$ if
$\Phi_{\bh} =$ constant, (eq.~\ref{eq:phi}; assuming $\dot{M}_{\rm
  fb} \propto t^{-\alpha}$) this limits the duration of Stage 1 to
$\sim 4^{-2/\alpha}t_{\rm MAD} \sim (1 /{\rm few})t_{\rm MAD}$, where $t_{\rm MAD}$ is
the onset of MAD accretion (start of Stage 2), which occurs when 
$\phi_\bh^{\rm MAD}\sim 50$.  The possibly
substantial duration of this early jet smothering phase is
illustrated by Figure \ref{fig:fluxwdc}, which shows the time
evolution of $\phi_{\bh}$ in the case of a tidally-disrupted WD.  In
this case, the disruption itself happens on a timescale of $\sim$ tens
of minutes, but Stage 1 begins only at $t\sim 10$ days.  Also note that depending on how fast
magnetic flux is accumulated, Stage 1 could be even briefer. If
$\phi_{\bh}\simmore \phi_\bh^{\rm MAD}$ at peak accretion, then the jet
will enter the `wobbling' stage described next essentially from its
onset.

\subsection{Stage 2: MAD Onset, Erratic Wobbling Jet (Fig.~\ref{fig:cartoon}b).} 
\label{sec:wobbling_jet}
As $\dot{M}$ decreases from its peak value, $\phi_{\bh} \propto \dot{M}^{-1/2}$
(eq.~[\ref{eq:phi}]) increases and the magnetic field becomes increasingly important dynamically. 
Once $\phi_{\bh}$ exceeds a critical value
$\phi_{\bh}^{\rm MAD} \sim 50$ (depending weakly on BH spin;
eq.~[\ref{eq:phiMAD}]), the field is sufficiently strong to
feed-back on the accreting gas, leading to a state of
`magnetically-arrested disk' accretion, or MAD (\citealt{nia03,tch11,mtb12}).

The strong magnetic flux also acts to orient the rotational axis of
the inner accretion disk/jet with the BH spin
axis \citepalias{mtb12b}. This realignment does not, however, happen
instantaneously, nor is it clean. The jets undergo a period of
vigorous rearrangement during which they ram against the accretion
disk, partially obliterate it, and intermittently punch holes through
the disk as they work to reorient the disk's angular momentum along BH
spin axis (see Fig.~\ref{fig:cartoon}b). Recent numerical simulations
show that during this stage the jets wobble intensely between the
orientation of the outer disk axis and the orientation of the BH axis
\citepalias{mtb12b}, so that jet emission transiently comes in and out of
our line of sight. This jet wobbling state may explain the epoch of intense flaring comprising the first $\sim10$ days of the \ubs~light curve ($\S\ref{sec:earlyflaring}$). 

Alignment between the disk/jet and BH spin completes once
sufficient magnetic flux is able to leak out of the immediate
vicinity of the BH and new flux is brought in by the infalling debris, such that the {\it entire} disk
(at least out to the circularization radius $R_{\rm circ}$) becomes
magnetically arrested. Depending on $R_{\rm circ}$, this process
requires $\phi_{\bh}$ to increase by a factor of a few, and hence $\dot{M}$
to decrease by a factor of several.  In $\S\ref{sec:flaring}$ we show that this timescale for the entire disk to `go MAD' is broadly consistent with the observed duration of the early flaring state in \ubs.  In Figures \ref{fig:fluxmsp} and \ref{fig:fluxwdc} this process is shown by the rising value of the disk flux $\Phi_{\rm D}$ during Stage 2, until it eventually reaches a constant fraction of the BH flux $\Phi_{\bh}$ at the onset of Stage 3.

\subsection{Stage 3: Steady Spin-Aligned Jet
  (Fig.~\ref{fig:cartoon}c).} 
\label{sec:stage-3:-steady}
Once the entire disk is magnetically-arrested, the system enters a scale-free MAD solution that depends on just one
parameter: the mass accretion rate (Fig.~\ref{fig:cartoon}c).\footnote{Order unity
  deviations from
  self-similarity are possible due to radiation feedback on the
  accretion flow.} The strong centrally-accumulated magnetic field not only aligns the jets along the direction of BH spin axis (\S\ref{sec:wobbling_jet}), it also explains
why the jet luminosity faithfully tracks mass fallback rate.  When
$\phi_{\bh}$ is small at early times (Stage 1), the BH flux $\Phi_{\bh}$ and jet
power $P_{j}$ (eq.~[\ref{eq:pj}]) are approximately constant or
increase in time (Stage 1), which is
inconsistent with the power-law decrease in the X-ray lightcurve of
\ubs\ between $\sim10$ and $\sim500$ days since the trigger
(Fig.~\ref{fig:Fx}). In contrast, when the central magnetic field is
sufficiently strong ($\phi_{\bh} \gtrsim \phi_{\bh}^{\rm MAD} \sim 50$;
eq.~[\ref{eq:phiMAD}]) to be dynamically-important (MAD state), then
the magnetic flux $\Phi_{\bh}$ threading the BH is not determined by initial
conditions, but instead by the ram pressure of the accretion flow. Its
value regulates such that the jet power $P_{\rm j} \sim
\eta_{j}\dot{M}$c$^{2}$ is a constant fraction $\eta_{j}
\approx 1.3a^{2}$ of the accretion power (eq.~[\ref{eq:Pjet}]). Thus,
as the fall-back rate decreases as a power-law in time, so do the jet
power and luminosity, as is observed.

Another expected feature of MAD is a stable QPO at a frequency that is
$1/4$ of BH horizon angular frequency \citep{mtb12}. \citet{reis12}
detected a potential QPO with period $\tau_{\rm QPO} \sim 200$ s in
the power-law decay portion of the \ubs~lightcurve, which could also
be evidence for MAD. We discuss the constraints implied by this period
($\S\ref{sec:qpo}$) as a part of our analysis in
$\S\ref{sec:constraints}$.

\subsection{Stage 4: No Jet (Fig.~\ref{fig:cartoon}d).} Once the
accretion rate drops below a few tens of per cent of $\dot{M}_{\rm
  edd}$, the flow transitions to a geometrically thin disk state, or
thermal state, which is not observed (nor theoretically expected) to
power a jet
\citep[e.g.,][]{fend04a,2011ApJ...739L..19R}.
The abrupt decrease in the X-ray flux at $t \sim 500$ days
(Fig.~\ref{fig:Fx}; by more than two orders of magnitude) can thus
plausibly be associated with the point at which $\dot{M} \sim
0.3\dot{M}_{\rm Edd}$ \citep{2009PASP..121.1279S,2010ApJ...718L.117S,2010A&A...521A..15A}. The jet luminosity and timescale of this transition also constrain the properties of the disrupted object and central BH ($\S\ref{sec:shutoff}$).

\subsection{Stage 5: Jet Revival (Fig.~\ref{fig:cartoon}e).} 
As $\dot{M}_{\rm fb}$ continues to decrease as a power-law in time,
eventually it will reach a few percent of
$\dot{M}_{\rm Edd}$. After this point the disk will again transition to a
geometrically-thick disk, analogous to the `low/hard' state
observed in X-ray binaries. Since this state is conducive to jet
formation \citep{nar95a,fend04a}, the jet in \ubs~and its
associated X-ray emission may suddenly turn back on (this is estimated to occur
sometime around 2016-2022; $\S\ref{sec:predictions}$).

\section{Constraints on \ubs}
\label{sec:constraints}

\subsection{Stellar Progenitor Scenarios}
\label{sec:star}
We begin by overviewing the possible classes of stellar progenitors
which could be responsible for \ubs. TD scenarios usually consider the
disruption of a lower main sequence star. However, in principle the
star could have been a giant \citep{MacLeod+12}, white dwarf
(e.g.~\citealt{Krolik&Piran11}; \citealt{Haas+12}; \citealt{Shcherbakov+12}), or even a planet.  Low mass planets seem to be ruled out because the
[beaming-corrected] energy budget of \ubs~of $\gtrsim$ few$\times
10^{51}$ erg already exceeds the rest mass energy of Jupiter.  A giant star is also unlikely, because the fall-back time of the stellar
debris would greatly exceed the observed trigger time ($\S\ref{sec:trigger}$), incompatible with the observed X-ray light
curve.  However, a white dwarf companion cannot be ruled out {\it a priori}, especially considering their potential for harboring a
large magnetic flux.

In what follows, we thus consider constraints based on two progenitor scenarios:  a lower main sequence star
(MS) and a white dwarf (WD).  We employ approximate mass-radius relations for each given by
\begin{align}
r_\st &= m_\st, &{\rm [MS]}  \nonumber \\ 
r_\st  &= 0.013(m_\st/0.6)^{-1/3}, &{\rm [WD]}
\label{eq:radii}
\end{align}
where $m_\st$ is in solar masses and $r_\st$ is in solar radii. Since the stellar radius must exceed
the tidal radius, only low mass black holes with $M_{\bh} \lesssim
10^{5}-10^{6}M_{\odot}$ are capable of disrupting a WD, depending on
WD mass and BH spin \citep{Kesden12}. For a centrally-concentrated MS star, the
disruption process can be either `full' or`partial' \citep{Guillochon&Ramirez-Ruiz12}, resulting in different predictions for the rate of fall-back accretion ($\S\ref{sec:trigger}$).

In what follows we thus consider three scenarios: (i) a complete or
(ii) partial tidal disruption of a lower mass main-sequence star by a
supermassive BH; or (iii) a complete disruption of a white dwarf by an
intermediate-mass BH.

\begin{figure}
\begin{center}
    \includegraphics[width=1\columnwidth]{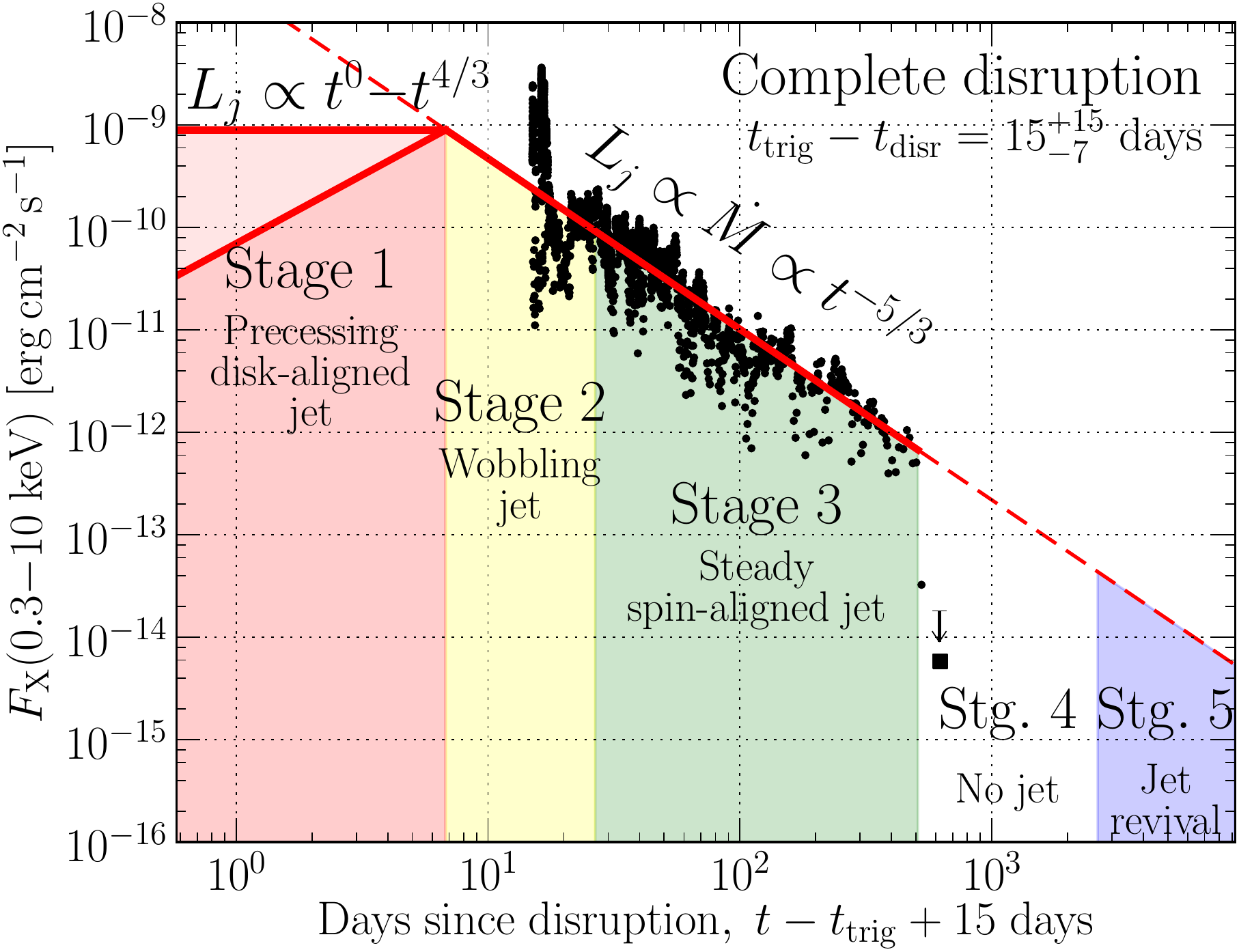}
  \end{center}
  \caption{\ubs's soft X-ray lightcurve accounting for the best-fit
    time of the disruption, $t_{\rm trig}-t_{\rm disr}=15$ days before the trigger, for a \emph{complete} disruption
    (see also Fig.~\ref{fig:Fx}). Note that just this simple shift in the zero
    point in time causes the early-time ``plateau'' in the light curve
    (see Fig.~\ref{fig:Fx}) to
    disappear and the entire light curve to become consistent with a
    single power-law dependence in time, $\propto t^{-5/3}$.
    Various stages of the disruption
    process are indicated with color coding and text. These stages are
    explained in Fig.~\ref{fig:cartoon} and its caption.}
  \label{fig:tdc}
\end{figure}

\begin{figure}
\begin{center}
    \includegraphics[width=1\columnwidth]{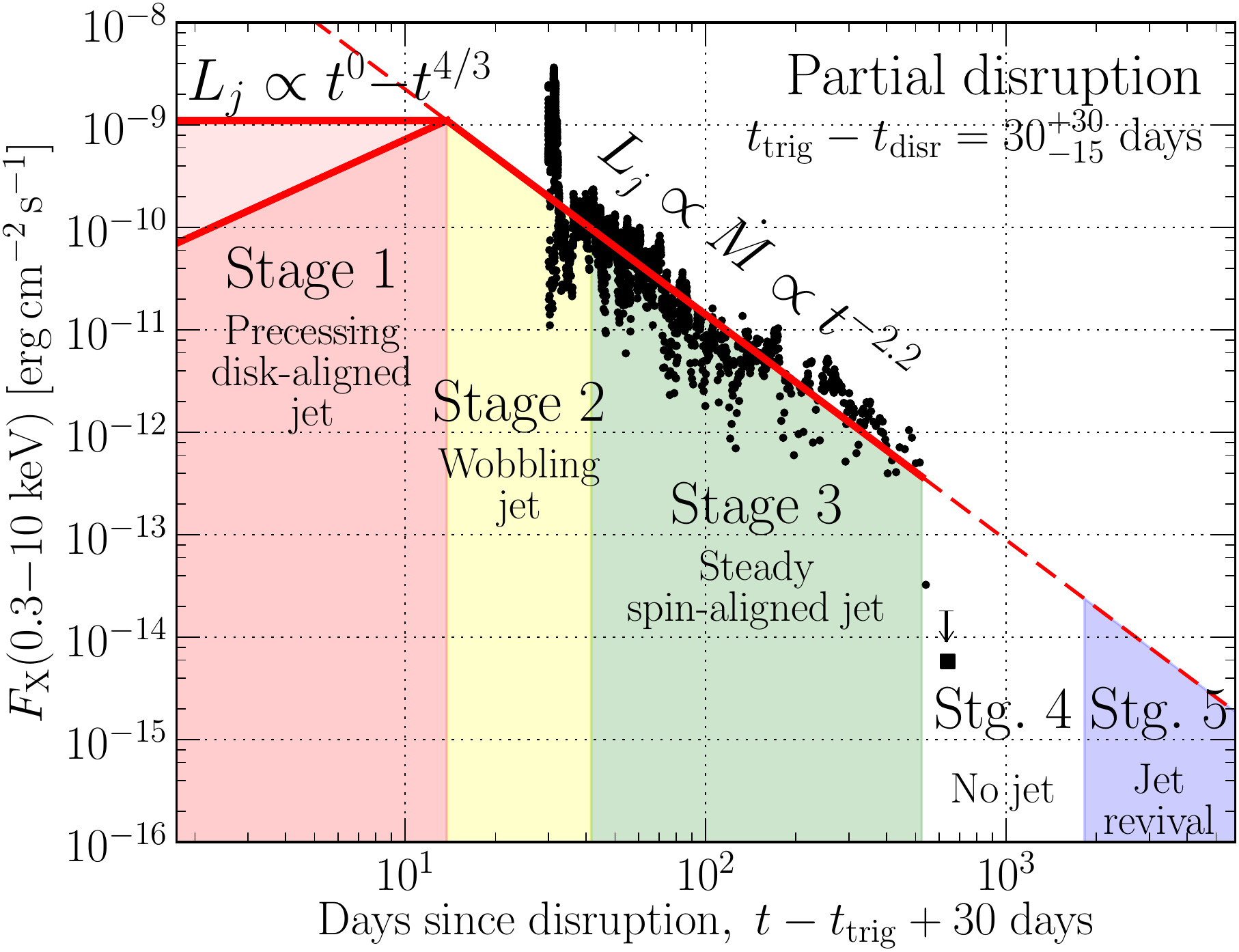}
  \end{center}
  \caption{\ubs's soft X-ray lightcurve accounting for the best-fit
    time of the disruption, $t_{\rm trig}-t_{\rm disr}=30$ days before
    the trigger, for a \emph{partial} disruption (see also
    Fig.~\ref{fig:Fx}). The steeper time-dependence of mass-fallback
    rate than in a complete disruption, $\propto t^{-2.2}$,
    allows for a longer delay between the time of the disruption and
    the time of the $\gamma$-ray trigger.}
  \label{fig:tdp}
\end{figure}

\subsection{Observational Constraints}
\label{sec:data}

For each progenitor scenario our goal is to determine the following 4 unknowns: BH mass,
$M_{\bh,5}$, BH spin, $a$, stellar mass, $m_\st$, and magnetic flux threading the hole, $\Phi_{\bh}$. We now discuss what observational data
constrain these properties within the framework of our model for \ubs.

\subsubsection{Shape of the X-Ray Light Curve}
\label{sec:trigger}

First consider what constraints can be placed on \ubs~based on the shape of the X-ray light curve.  Recall that although the time-averaged light curve does not follow a single power-law decay if plotted against the time since the $\gamma-$ray trigger $t - t_{\rm trig}$, a much better fit is achieved by moving the TD `zero point' prior to the trigger (Fig.~\ref{fig:Fx}).  We thus first consider what range of trigger delay time $t_{\rm trig} - t_{\rm disr}$ produces a light curve shape consistent with the predicted power-law decline in the X-ray luminosity $L_{X} \propto \dot{M}_{\rm fb} \propto (t-t_{\rm disr})^{-\alpha}$ given the expected range in $\alpha$.

Although the canonical value of $\alpha = 5/3$ is often quoted, $\alpha$ actually depends on the fraction of mass lost by the star during TD, $\Delta M_\st/M_\st=0.4f_{0.4}$, which differs between complete and partial disruptions (\citealt{Guillochon&Ramirez-Ruiz12}).  If $\Delta M_{\st}/M_{\st} \approx 10\%-50\%$ then $\alpha=2.2$, but if $\Delta M_{\st}/M_{\st} \gtrsim 50\%$ then the value of $\alpha$ approaches that for a complete disruption, $\alpha=5/3$ (\citealt{Guillochon&Ramirez-Ruiz12}).  By considering the two limiting cases of complete disruptions with $\alpha=5/3$ and partial disruptions with $\alpha=2.2$, we bracket the allowed range of possibilities.
 
For complete disruptions ($\alpha = 5/3$) we find an allowed range of $t_{\rm trig}-t_{\rm disr}\sim 15^{+15}_{-7}$~days when fitting to the shape of the time-averaged X-ray light curve.  Figure \ref{fig:tdc} shows the trigger delay-corrected light curve for a fiducial value $t_{\rm trig}-t_{\rm disr} = 15$ days.  For partial disruptions ($\alpha = 2.2$), we find required delay times that are somewhat longer $t_{\rm trig}-t_{\rm disr}\sim 30^{+30}_{-15}$~days.  Figure~\ref{fig:tdp} shows the corrected light curve in this case for a fiducial delay $t_{\rm trig}-t_{\rm disr} = 30$~days.

Our fits quoted above (Figs.~\ref{fig:tdc}, \ref{fig:tdp}) were made by matching the fall-back accretion rate $\propto$ jet power to the time-averaged X-ray flux.  However, if the early phase of high variability indeeds results from a `wobbling' jet ($\S\ref{sec:wobbling_jet}$), then bright flares may arise from transient episodes when the jets points towards our line of site.  In this case, it is more physical to match the fall-back rate $\propto$ jet power to the `envelope' in the X-ray light curve comprised by the flare peaks, since these better characterize the true jet power.   Since the observed duty cycle of the flaring state is $\sim 10\%$, this implies that the jet points towards us only about $1/10$ of the time if the wobbling scenario is correct.  The total jet power [accounting for both on- and off-axis jet emission] could thus exceed the observed [on-axis] emission by an order of magnitude (see also \S\ref{sec:radio}).  
Figure \ref{fig:Fx} shows that such fits allows for a trigger delay $t_{\rm trig}-t_{\rm disr}$ as short as a few days, consistent with the first evidence for a jet in \ubs~$\sim$ 4 days prior to the $\gamma-$ray trigger (\citealt{Krimm+11}).

Given the allowed range in $t_{\rm trig}$, we require that the fallback accretion rate must be past its peak at the trigger:
\begin{equation}
  t_{\rm peak} = 1.5t_{\rm fb} < t_{\rm trig}/(1+z),
  \label{eq:tpeakcond}
\end{equation} where the fall-back time (eq.~[\ref{eq:tfallback}]) specialized to WD and MS scenarios is given by
\begin{subequations}     \label{eq:tfb}
\begin{align}
  \label{eq:tfbwd}
  t_{\rm fb} &= 0.02\,{\rm d}\; M_{\bh,5}^{1/2} m_\st^{-3/2}, &{\rm [WD]}\\
    t_{\rm fb} &= 17.3\,{\rm d}\; M_{\bh,5}^{1/2} m_\st^{1/2}, &{\rm [MS]},
\end{align}
\end{subequations}
since otherwise the light curve would still be rising and hence would not match the $L_X \propto \dot{M}_{\rm fb} \propto (t/t_{\rm fb})^{-\alpha}$ decay predicted for $t \gg t_{\rm fb}$.

\subsubsection{Magnetic Flux} 
Although it does not represent an independent constraint, the magnetic flux $\Phi_{\bh}$ is determined once the spin of the BH is known, if one assumes that the jet is in a MAD state after the trigger, up until the point when the jet shuts off.  During MAD accretion, the BH magnetic flux is regulated by the mass accretion rate to a dimensionless value \citep{tmn12,tm12b}, which is well approximated as
\begin{equation} \label{eq:phiMAD} 
\operatorname{In\ a\ MAD:}\quad \phi_{\bh}^{\rm MAD} \equiv \frac{\Phi_{\bh}^{\rm MAD}}{(\dot
    M r_g^2c)^{1/2}} \approx 70(1-0.38\omega_{\rm H})\hr^{1/2},
\end{equation}
corresponding to absolute flux
\begin{equation}
  \label{eq:Phi30}
\operatorname{In\ a\ MAD:}\quad  \Phi_{\bh,30}^{\rm MAD} = 0.067 M_{\bh,5}^{3/2}\lambda^{1/2}(1-0.38\omega_{\rm H}) \hr^{1/2},
\end{equation}
where $\lambda=\dot M/\dot M_{\rm Edd}$ is the Eddington ratio and
$h/r=0.3\hr$ is the thickness of the accretion flow.  Thus, if
the dimensionless BH spin and the jet luminosity at a given Eddington
ratio are known (such as at the point of jet shut-off; see
eq.~[\ref{eq:mcr}] below), then $\Phi_{\bh}$ can be determined.

\subsubsection{Jet Shut-Off Power}
\label{sec:shutoff}

Another constraint is that the time at which the jet was observed to shut-off $t_{\rm off}-t_{\rm trig}\approx500$ days (see Fig.~\ref{fig:Fx})
happens simultaneously with the expected state transition occurring at a fraction of the Eddington accretion rate (\S\ref{sec:jet-shutoff}), i.e.~
\begin{equation}
  \label{eq:mcr}
  \dot M_{\rm cr} = 0.3\frac{\lambda_{\rm cr}}{0.3} \dot M_{\rm Edd},
\end{equation}
where theoretical and observational uncertainties place the threshold value in the range $0.1\lesssim \lambda_{\rm
  cr}\lesssim0.5$.

Just prior to when the jet shut off, its X-ray luminosity $\propto$ jet power $P_{\rm j}^{\rm off}$ was $\sim 200$ times smaller than its value at the end of the $\sim 10$ day plateau phase after the trigger, \begin{equation}
  \label{eq:pjoff}
  P_{\rm  j}^{\rm off} \simeq (1/200) P^{\rm trig}_{\rm j}
  \approx5\times10^{43} \frac{f_{\rm
  b}\epsilon_{\rm bol}\epsilon_{X}^{-1}}{0.03}\ {\rm erg\,s^{-1}}.
\end{equation}
Now, by combining equations \eqref{eq:pj} and \eqref{eq:phiMAD}, the jet power can be written:
\begin{align}
  \label{eq:Pjet}
\operatorname{In\ a\ MAD:} \quad
P_{\rm j} 
&\approx F(\omega_{\rm H}) \hr\dot M c^2 \approx 1.3 \hr a^2\dot Mc^2 \notag\\
&\approx 1.6\times 10^{44} \hr\lambda  M_{\bh,5}\; a^2  \ {\rm
    erg\,s^{-1}},
\end{align}
where $\lambda \equiv \dot{M}/\dot{M}_{\rm Edd}$ and we used the fact
that the spin-dependent factor entering jet power, $F(\omega_{\rm
  H})=4.4 \omega_{\rm H}^2 (1-0.38 \omega_{\rm H})^2 f(\omega_{\rm
  H})$, can be approximated as $F\approx 1.3 a^2$ to $10$\% accuracy
for $0.3\le a \le1$ \citep{tm12b}.

Matching the jet power with the observed power (eq.~[\ref{eq:pjoff}]) at $\dot M=\dot M_{\rm cr}$
(eq.~[\ref{eq:mcr}]) thus requires
\begin{equation}
  \label{eq:pjet_condition}
a^2 M_{\bh,5} = \frac{f_{\rm
  b}\epsilon_{\rm bol}\epsilon_{X}^{-1}}{0.03}\frac{0.3}{\lambda_{\rm
  cr}} \frac{0.3}{h/r}.
\end{equation}
This constraint is independent of the nature of the disrupted object.

\subsubsection{Jet Shut-Off Accretion Rate}

Another constraint is that  the BH mass accretion rate must equal
the critical accretion rate (eq.~[\ref{eq:mcr}]) at the observed jet shutoff time, $t_{\rm off}-t_{\rm trig}\simeq500$~days, viz.~
\begin{equation}
\label{eq:mdotoff}
\dot M(t_{\rm  off}) = \dot M_{\rm cr},
\end{equation}
where $\dot{M} = f_{\rm acc}\dot{M}_{\rm fb}$ and the predicted fall-back accretion rate (eqs.~[\ref{eq:tfallback}]-[\ref{eq:mdotpeak}]) specialized to the MS and WD scenarios:
\begin{align}
  \dot{M}_{\rm fb} &= 4.2\times10^{28} M_{\bh,5}^{1/3}m_{\st}^{4/3}t_{1}^{-5/3}{\rm g\;s^{-1}},\quad{\rm [MS,\ complete]} \\
  \dot M_{\rm fb} &= 1.2\times 10^{29} f_{0.4} 
  M_{\bh,5}^{3/5}m_\st^{8/5}t_{1}^{-2.2}\ {\rm g\;s^{-1}},\quad{\rm [MS,\ partial]} \label{eq:mfbmsp}\\
\dot{M}_{\rm fb} &= 4.6\times10^{26} M_{\bh,5}^{1/3}t_{1}^{-5/3}{\rm
  g\;s^{-1}},\quad{\rm [WD,\ complete]}
\label{eq:mfbwdc}
\end{align} 
where $t_{1} \equiv t/$day.  

The factor $f_{\rm acc} < 1$ accounts for the possibility that only a
fraction of the fall-back material actually reaches the BH horizon,
with the rest either expelled in the form of the accretion disk winds
and or lost during the circularization of the tidal streams.
Super-Eddington accretion is susceptible to outflows driven by
radiation pressure (e.g.~\citealt{Ohsuga+05}, \citealt{Strubbe&Quataert09}), 
but the magnitude of this effect is uncertain. 
Fortunately,  we find that the lower limit on $f_{\rm acc}$ is not constraining (\S\ref{sec:results}), 
so without the loss of generality we allow the full range $0 < f_{\rm acc}\le1$ in our calculations.

\subsubsection{Spin Sufficient for Alignment}
\label{sec:spinconstraint}

We require that the spin of the BH to be sufficiently high, 
\be
a\gtrsim0.5,
\label{eq:spin_condition}
\ee
such that the BH is able to magnetically align the disk and the jets \citepalias{mtb12b}. 

\subsubsection{MAD QPO}
\label{sec:qpo}

A final possible constraint is to match the QPO period measured in the power-law X-ray light curve of \ubs\ \citep{reis12}
\begin{equation}
  \label{eq:tqpo}
  \tau_{\rm QPO} = 210\pm30\ {\rm s}.
\end{equation}
with the predicted MAD QPO, which occurs at a frequency that is
$1/4$ of BH horizon angular frequency \citep{mtb12}, 
$\Omega_{\rm H} = ac/(2r_{\rm H})=\omega_{\rm H}c/(2r_{\rm g})$, resulting in a predicted period
\begin{equation}
  \label{eq:taumad}
  \tau_{\rm MAD} = \frac{2\pi}{0.25\Omega_{\rm H}} = \frac{16
    \pi}{\omega_{\rm H}} \frac{r_{\rm g}}{c} = 24.8\, M_{\bh,5}\, \omega_{\rm
    H}^{-1}\ {\rm s}.
\end{equation} 
Equating \eqref{eq:taumad}, multiplied by $(1+z)\approx1.353$, with \eqref{eq:tqpo} gives a final constraint,
\begin{equation}
  \label{eq:qpocond}
    M_{\bh,5} = (6.3\pm0.9) \omega_{\rm H}.
\end{equation}
We show below that this requirement is constraining only in the WD scenario.

\begin{figure}
\begin{center}
    \includegraphics[width=0.8\columnwidth]{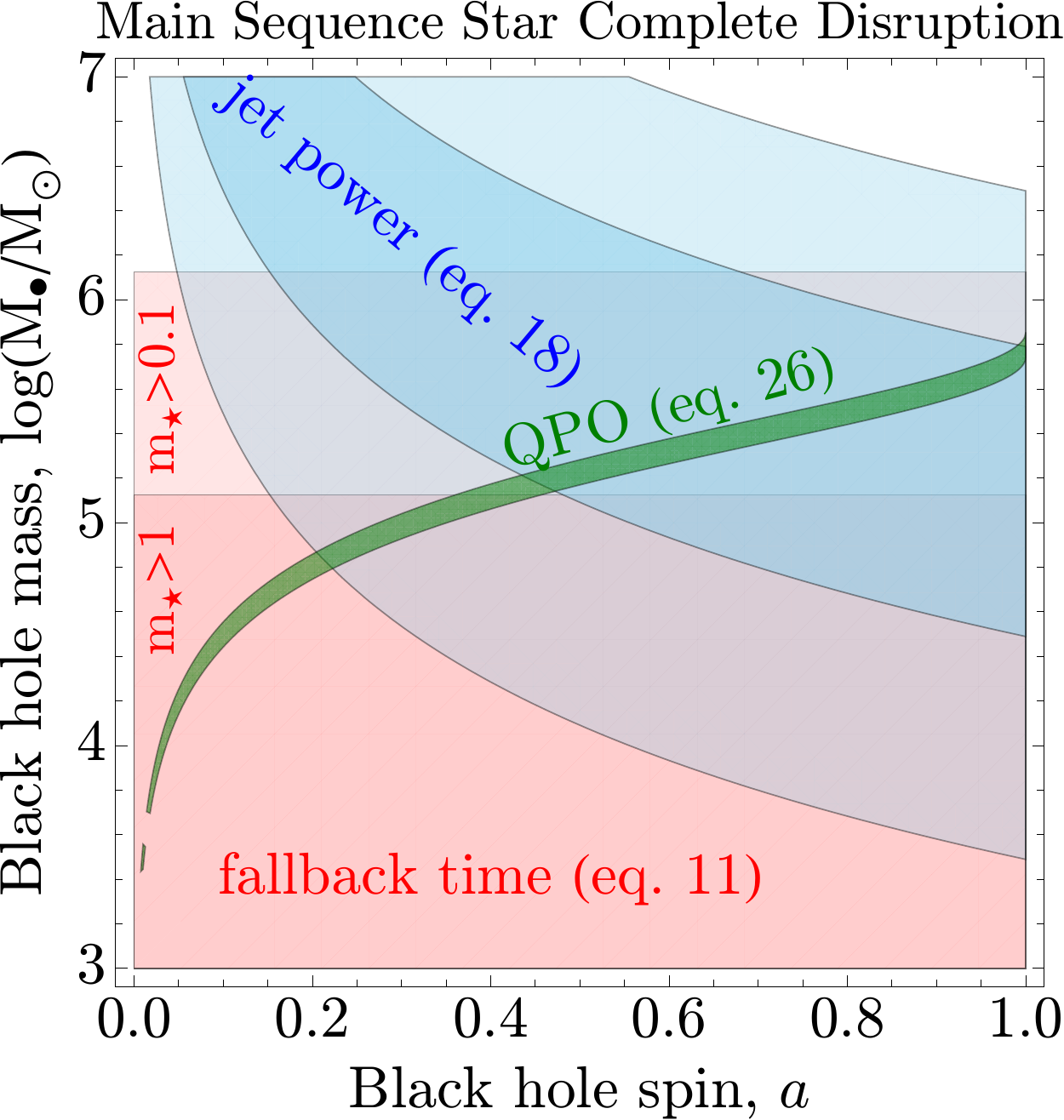}
  \end{center}
  \caption{Constraints on BH mass, $M_\bh$, and spin, $a$, in
    the scenario of a complete disruption of a lower-mass
    main-sequence star. The constraint on jet power at jet shut off [eq.~\eqref{eq:pjet_condition}]
    gives the blue-shaded regions, with the darker shaded region more
    likely. The constraint on disruption timescale
    [eq.~\eqref{eq:tpeakcond}] gives the dark (light) shaded red
    regions for $m_\st>1$ ($m_\st>0.1$, respectively). The QPO period
    constraint [eq.~\eqref{eq:qpocond}] gives the green curve. The
    common region seems to favor a lower-mass star,
    $m_\st\sim{\rm few}\times0.1$, and a light BH, $M_{\bh,5} \sim {\rm few}$, with $a\gtrsim0.5$. 
  }
  \label{fig:msf}
\end{figure}

\subsection{Results}
\label{sec:results}
Equations (\ref{eq:tpeakcond}), (\ref{eq:pjet_condition}),
(\ref{eq:mdotoff}), (\ref{eq:spin_condition}) and (\ref{eq:qpocond})
provide 4 or 5 constraints on the unknown parameter space ($M_\st,
M_{\bh}, a, $), depending on whether one adopts the (speculative)
QPO constraint described in $\S\ref{sec:qpo}$. Once $M_{\bh}$
and $a$ are determined, the magnetic flux $\Phi_{\bh}$ follows from equation
(\ref{eq:Phi30}). We now apply these constraints individually to each
of our proposed scenarios ($\S\ref{sec:star}$): Complete Disruption of
a Main Sequence Star ($\S\ref{sec:compl-disr-main}$); Partial
Disruption of a Main Sequence Star ($\S\ref{sec:part-disr-main}$); and
Complete Disruption of a White Dwarf ($\S\ref{sec:compl-disr-white}$).

\subsubsection{Complete Disruption of a Main-Sequence Star}
\label{sec:compl-disr-main}

Figure \ref{fig:msf} summarizes the constraints on the BH mass and spin for the complete tidal disruption of a low mass MS star.  The first constraint, based on the shape of the X-ray light curve (eq.~[\ref{eq:tpeakcond}]), can be written
\begin{equation}
  \label{eq:mscpeakcond}
  M_{\bh,5} < 1.3 m_\st^{-1},
\end{equation}
where we have adopted the highest value for $t_{\rm trig}=30$~days that allows us to reproduce the observed shape of the X-ray lightcurve (Figs.~\ref{fig:Fx}, \ref{fig:tdc}).  This constraint is shown in Figure \ref{fig:msf} as the dark (light) red shaded regions for $m_\st \ge 1$ ($m_\st \ge
0.1$). 

For a solar mass star, equation (\ref{eq:mscpeakcond}) places a tight
upper limit on the BH mass, $M_\bh < 1.3\times10^5M_\sun$. Such a low
mass BH would be novel since they are quite rare (either
intrinsically, or due to the observational challenges in detecting
them; e.g.~\citealt{Greene12}) and might be unexpected given the host
galaxy of \ubs. A higher mass $M_\bh\gtrsim {\rm few}\times10^5M_\sun$
is allowed if the mass of the star is lower, $m_\st \lesssim 0.5$,
below the peak of the standard IMF \citep{Chabrier03}. 

The second constraint (eq.~[\ref{eq:pjet_condition}]) cuts out an stripe in the $M_\bh{-}a$ plane, as shown as a blue-colored region in Figure \ref{fig:msf}. The width of the dark (light) blue region reflect an optimistic (conservative) factor of 20 (1000) uncertainty in the value of the right-hand side of equation \eqref{eq:pjet_condition}.  The chief effect of this constraint is to place a lower limit on the BH mass and spin, the latter consistent with the fourth constraint ($\S\ref{sec:spinconstraint}$).

The third constraint (eq.~[\ref{eq:mdotoff}]) does not place an interesting limits on BH mass or spin due to the uncertainty in the fraction of the stellar material accreted $f_{\rm acc}$.  However, it does pick out a preferred range in the value of $f_{\rm acc}$ given the other parameters:
\begin{equation}
f_{\rm acc} = 0.02 \frac{\lambda_{\rm cr}}{0.3}M_{\bh,5}^{2/3}m_\st^{-4/3} < 0.024 \frac{\lambda_{\rm cr}}{0.3}m_\st^{-2},
\label{eq:faccmsf}
\end{equation}
where in the inequality we have used equation (\ref{eq:mscpeakcond}).  Equation \eqref{eq:faccmsf} shows that if the tidally disrupted star was solar-like ($m_\st \sim 1$), then large mass loss is required, as could be the result of outflows from the disk or at the impact point of tidal streams.   Alternatively, a higher value $f_{\rm acc} \sim 1$ is allowed if the progenitor is instead a low mass M star $m_\star \sim 0.1-0.2M_{\odot}$ near the hydrogen-burning limit. 

The fifth constraint, on the QPO frequency, produces an allowed region shown with green in Fig.~\ref{fig:msf}, which is consistent with (and does not appreciably alter) our conclusions above.  If taken seriously, this constraint places an upper limit on the BH mass $M_\bh<6\times10^5M_\sun$, but otherwise a wide range of BH spin, $a\gtrsim0.5$, is allowed.

\begin{figure}
\begin{center}
    \includegraphics[width=0.8\columnwidth]{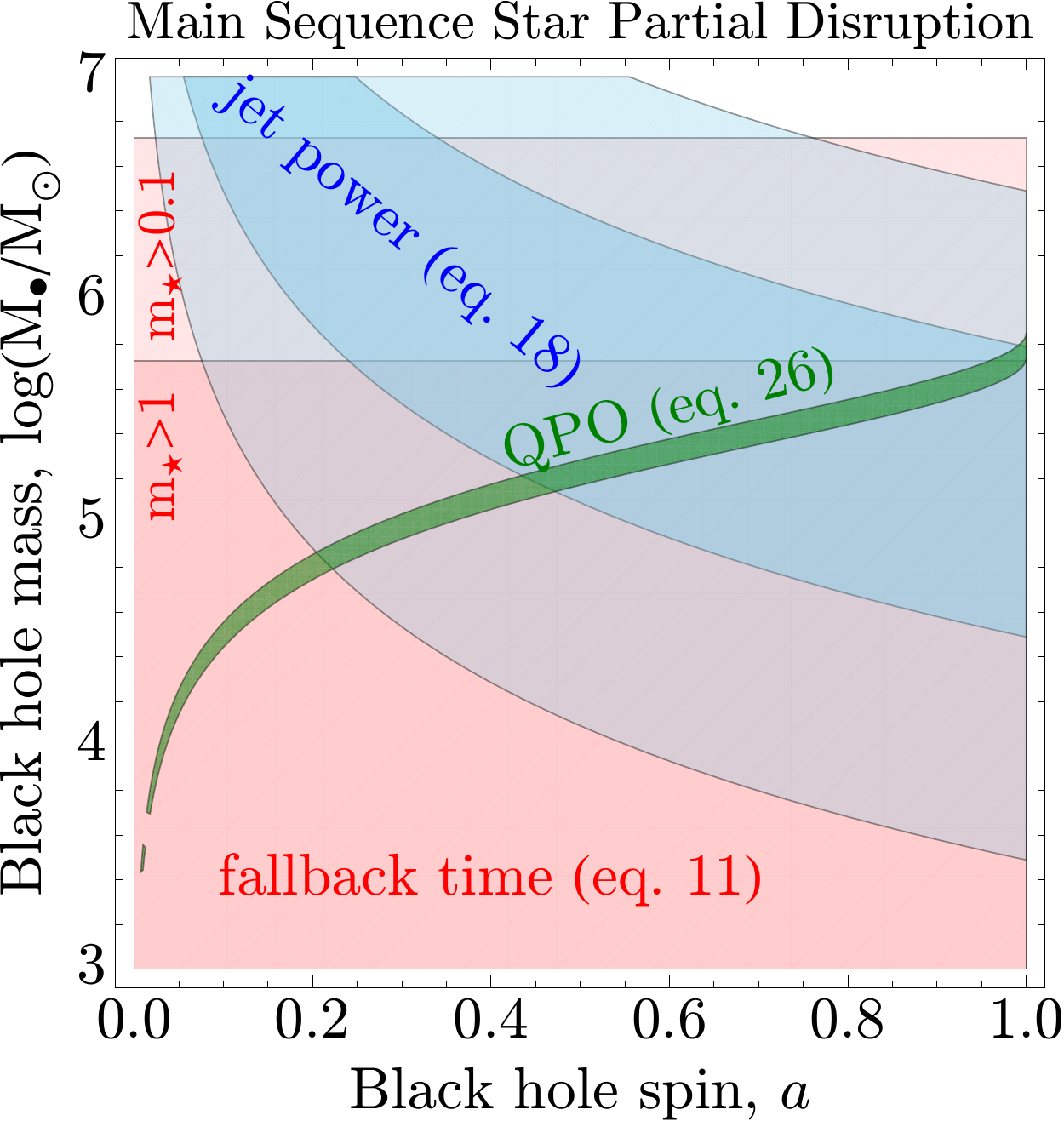}
  \end{center}
  \caption{Constraints on BH mass, $M_\bh$, and spin, $a$, in
    the scenario of a partial disruption of a lower-mass
    main-sequence star. The constraint on jet power at jet shut off [eq.~\eqref{eq:pjet_condition}]
    gives the blue-shaded regions, with the darker shaded region more
    likely. The constraint on disruption timescale
    [eq.~\eqref{eq:tpeakcond}] gives the dark (light) shaded red
    regions for $m_\st>1$ ($m_\st>0.1$, respectively). The QPO period
    constraint [eq.~\eqref{eq:qpocond}] gives the green curve. The
    common region (including the QPO constraint) 
    allows a Sun-like star, $m_\st\lesssim1$, and a 
    super-massive BH, $M_{\bh,5}\lesssim5$, with $a\gtrsim0.5$.}
  \label{fig:msp}
\end{figure}

\subsubsection{Partial Disruption of a Main-sequence Star}
\label{sec:part-disr-main}
Figure \ref{fig:msp} summarizes the constraints on the BH mass and spin based on our second scenario, the partial tidal disruption of a lower-mass MS star.  As in equation (\ref{eq:mscpeakcond}), the first constraint can be written
\begin{equation}
  \label{eq:msppeakcond}
  M_{\bh,5} < 5.3 m_\st^{-1},
\end{equation}
where here we have taken $t_{\rm trig}=60$~days (again as the maximum allowed by fitting the observed shape of the light curve; \S\ref{sec:compl-disr-main}). 
Constraint (\ref{eq:msppeakcond}) is a factor of several less restrictive than that for a complete disruption (eq.~\ref{eq:mscpeakcond}).  For example, the data now allow a relatively massive BH with $M_\bh\gtrsim{\rm few}\times10^5M_\sun$, even for a Sun-like star, $m_\st\simeq1$. 

The region allowed by the second constraint (eq.~[\ref{eq:pjet_condition}]; again shown in blue) is exactly the same as for the full disruption (Fig.~\ref{fig:msf}).  The QPO condition is again only moderately constraining, placing an upper limit $M_{\bh,5}<6\times10^5M_\sun$ on the BH mass.

Similar to the full disruption, the third constraint does not place any interesting limits on the BH or stellar parameters, but it does tell us the fraction,
\begin{equation}
f_{\rm acc} = 0.2 f_{0.4} \frac{\lambda_{\rm cr}}{0.3}M_{\bh,5}^{2/5}m_\st^{-8/5} < 0.39 f_{0.4} \frac{\lambda_{\rm cr}}{0.3}m_\st^{-2}.
\label{eq:faccmsp}
\end{equation}
of the fallback material reaching the BH, where equation (\ref{eq:msppeakcond}) is used in the last inequality.  For low mass stars the required value of $f_{\rm acc}$ may even exceed unity, potentially ruling out such progenitors depending on the value of $\lambda_{\rm cr} \sim 0.1-0.5$.  This behavior is opposite to that in the complete MS disruption scenario, where we were forced to conclude that $f_{\rm acc}\ll1$ [eq.~(\ref{eq:faccmsf})].  As we discussed in \S\ref{sec:trigger}, the two extreme possibilities--of a complete and partial disruption--bracket a continuous family of allowed solutions that, by continuity, are consistent with the data.

\begin{figure}
\begin{center}
    \includegraphics[width=0.8\columnwidth]{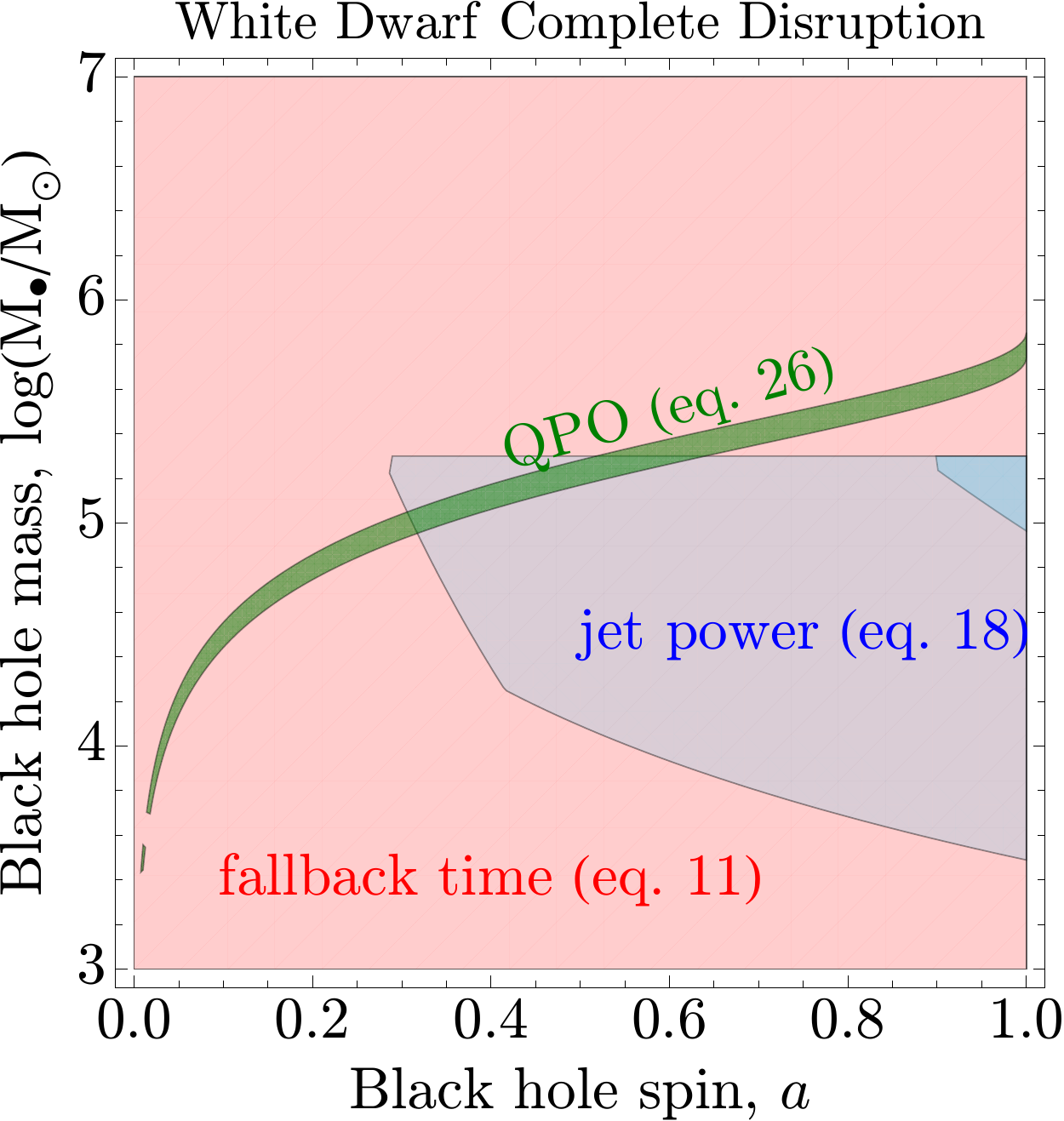}
  \end{center}
  \caption{Constraints on BH mass, $M_\bh$, and spin, $a$, in
    the scenario of a complete disruption of a white dwarf. 
    The constraint on jet power at jet shut off [eq.~\eqref{eq:pjet_condition}]
    gives the blue-shaded regions, with the darker shaded region more
    likely. This darker shaded region appears marginally within the bounds of this
    plot and requires extreme BH spin: this is due to
    limited mass supply of the WD that has hard time to power the jets
    as strong as they are observed. The constraint on disruption timescale
    [eq.~\eqref{eq:tpeakcond}] allows all of the parameter space. 
    We conclude that WD disruption scenario is possible for
    intermediate mass BHs, with $M_{\bh,5}=0.1{-}2$ and $a\gtrsim0.5$.
    In addition to requiring an intermediate-mass BH, the WD scenario also requires very small jet opening angles and small bolometric
    correction, $f_{\rm
      b}\epsilon_{\rm bol}<5\times10^{-4}$. Thus, we suggest that the
    WD disruption scenario is less likely than a MS star disruption.
    Accounting for the QPO period
    constraint [eq.~\eqref{eq:qpocond}], which is shown with the green curve,
    constrains the BH spin to $a\sim0.5$.}
  \label{fig:wdf}
\end{figure}

\subsubsection{Complete Disruption of a White Dwarf}
\label{sec:compl-disr-white}

Figure \ref{fig:wdf} summarizes the constraints on the BH properties based on our third scenario, the disruption of a WD.  Since a WD is much denser than a MS star, the fallback time in the WD scenario is much shorter than in the MS
scenario. In fact, $t_{\rm fb}\ll t_{\rm trig}$, where $t_{\rm trig} \gtrsim 5$~days is required to explain the shape of the X-ray lightcurve to within a factor of
few (see \S\ref{sec:trigger}). For these reasons, the first constraint, (\ref{eq:tpeakcond}), does not place interesting limits on the system parameters.

The most important constraint is that on the accretion rate, eq.\ (\ref{eq:mdotoff}), which
gives
\begin{equation}
  \label{eq:faccwd}
  f_{\rm acc} = 2 \frac{\lambda_{\rm cr}}{0.3} M_{\bh,5}^{2/3}.
\end{equation}
The physical requirement that $f_{\rm acc}<1$ places an upper limit on the BH mass,
$M_{\bh,5}\lesssim2$ (for $\lambda_{\rm cr} > 0.1$), thus {\it requiring an intermediate-mass BH}.

Since the mass fallback rate is independent of the WD mass (eq.~[\ref{eq:mfbwdc}]), this
makes the second constraint, eq.~(\ref{eq:pjet_condition}), on jet power
particularly constraining on the BH spin,
\begin{equation}
  \label{eq:amin}
  a=1.6 \left(\frac{f_{\rm
  b}\epsilon_{\rm bol}\epsilon_{X}^{-1}}{0.03}\right)^{1/2}
  \left(\frac{\lambda_{\rm cr}}{0.3}\right)^{1/4} 
  f_{\rm acc}^{-3/4}\hr^{-1/2}.
\end{equation}
The constraints resulting from eqs.~\eqref{eq:faccwd} and \eqref{eq:amin} 
are illustrated in Figure~\ref{fig:wdf} with the dark (light) blue area, whose width
reflects optimistic (conservative) modeling and observational
uncertainties of a factor $20$ ($100$) in the right-hand side of
eq.~\eqref{eq:pjet_condition}. The most likely (dark blue region)
favors a high BH spin, $a\gtrsim0.9$.

The addition of the less certain QPO constraint (eq.~\ref{eq:qpocond}) in combination with the spin constrain (eq.~\ref{eq:spin_condition}) favors a BH mass in the range $M_{\bh, 5} \sim 0.6-2$ with an intermediate spin $a\simeq0.5$.  However, we caution against over-interpreting the QPO constraint, as the observed QPO could be caused by other processes than those resulting from MAD accretion \citep{mtb12}.

Finally, we note that \emph{partial} disruption of a WD is less plausible than a complete disruption due to the lower mass fall-back rate and the steeper time-dependence of $\dot M_{\rm fb}$ (eq.~[\ref{eq:mfbwdc}]).  This makes constraint \eqref{eq:amin} much more severe, causing tension with the observations.  For this reason, we focus exclusively on complete disruption in the WD case.

\section{Discussion}
\label{sec:discussion}

\subsection{Nature of the Disrupted Star}
\label{sec:progenitor}

In $\S\ref{sec:star}$ we introduced three plausible scenarios for the nature of the disrupted object: a MS star (full or partial disruption) and a WD (full disruption).  We find that a MS stellar disruption comfortably satisfies all of the observational constraints (\S\ref{sec:data}; Figs.~\ref{fig:msf}, \ref{fig:msp}) for a BH with mass $M_\bh\sim({\rm few}{-}10)\times10^5M_\sun$ which is consistent with upper limits based on X-ray variability and the host galaxy of \ubs (\citealt{Levan+11}; \citealt{Bloom+11}; \citealt{Burrows+11}).  Both complete and partial disruption are allowed.  A complete disruption favors lower-mass stars, $M_\star\lesssim 0.5M_\sun$, while a partial disruption (in which a star loses $10$ to $50$\% of its mass while its core survives) allows a wide range of stellar masses, $M_\star=(0.1{-}1)M_\sun$.  Complete disruptions of more massive stars ($M_\star\sim M_\sun$) are also allowed but require (potentially rare) smaller-mass BHs ($M_\bh \lesssim 10^5M_\sun$).  

Unlike a MS star disruption, WD disruptions require the central object to be an intermediate-mass BH of mass $M_\bh = (1-10)\times10^4M_\sun$ (\S\ref{sec:compl-disr-white}; Fig.~\ref{fig:wdf}).  In addition, the mass fallback rate in the WD case is just barely sufficient to power the observed jets, which thus greatly restricts the allowed parameter space (\S\ref{sec:compl-disr-white}).

Although the WD scenario is formally allowed, there are several reasons to favor the MS scenario.  In addition to the narrow allowed parameter space in the WD case, the probability of WD disruption is much lower than in the MS since its smaller tidal radius limits the allowed range of impact parameters capable of resulting in disruption.  Furthermore, the intermediate-mass BH required in the WD case is an exotic, and probably rare, object (\citealt{Greene12}); it is also unclear how frequently such an object would be expected to reside near the nucleus of a galaxy.  Given the convergence of several rare events in the WD scenario, we conclude that a MS disruption is more likely.  The existence of plausible and constrained solutions provides a consistency check on our model for \ubs.

In addition to jetted emission, TD events are accompanied by thermal optical/UV/soft X-ray emission from the accretion disk (e.g.~\citealt{Ulmer1999}) or super-Eddington outflows \citep{Strubbe&Quataert09,Strubbe&Quataert11}.  No such optical/UV emission was detected from \ubs, possibly due to substantial dust extinction (\citealt{Bloom+11}).  However, if an otherwise similar event with less extinction were to occur in the future, a measurement of the disk flux just after the jet shuts off would directly determine $\lambda_{\rm  cr}$, the critical Eddington ratio at which the disk transitions from being thick to thin.  Combined with the jet-related constraints in $\S\ref{sec:data}$, such a measurement would also better determine the properties of the BH and disrupted star.

\subsection{Origin of the Magnetic Flux}
\label{sec:flux}

Regardless of the nature of the disrupted star, the magnetic flux threading the BH must be sufficient to power jet responsible for~\ubs. Equation \eqref{eq:pjPhi30} can be written as:
\begin{eqnarray}
  \label{eq:Phi30sol}
  \Phi_{\bh,30}^{\rm trig} \approx 0.4 M_{\bh,5} \left(\frac{P_{\rm j}^{\rm trig}}{10^{46}\
      {\rm erg\;s^{-1}}}\right)^{1/2} \;\left(\frac{\omega_{\rm H}f^{1/2}(\omega_{\rm H})}{0.64}\right)^{-1}
\end{eqnarray}
where $\omega_{\rm H}f^{1/2}(\omega_{\rm H})$ is normalized to its value for BH spin $a = 0.9$.  If the onset of MAD indeed occurred near the time of the $\gamma-$ray trigger  (at Eddington ratio $\lambda_{\rm trig}\approx200\lambda_{\rm cr}\approx60(\lambda_{\rm cr}/0.3)$; see~\S\ref{sec:shutoff}), then we require $\Phi_{\bh,30}^{\rm MAD}\sim 0.1-10$ for BH masses $M_{\bh, 5} \sim 0.1-10$ consistent with our modeling of \ubs~(Figs.~\ref{fig:msf}-\ref{fig:wdf}).  This flux is $\sim 4-6$ orders of magnitude greater than that through a solar-type star $\Phi_{\star} \sim \pi B_{\star}R_{\star}^{2} = 10^{25}(B_{\star}/$kG$)(R_{\star}/R_{\odot})^{2}$ G cm$^{2}$, even for an optimistically large $B_{\star} \sim $ kG average stellar magnetic field.  Likewise, a WD ($R_{\star} \sim 0.01R_{\odot}$) would require a field $B_{\star} \gtrsim 10^{11}$ G for $M_{\bh, 5} \sim 1$ (Fig.~\ref{fig:wdf}) which exceeds the largest measured [surface] fields by two orders of magnitude (e.g.~\citealt{Kepler+12}).   

If the star itself cannot explain the flux, what could be its source?  It has been suggested \citep{Krolik&Piran11,Krolik&Piran12} that the requisite flux is generated by a turbulent dynamo in the accretion flow.  However, since the {\it net} vertical magnetic flux is a conserved quantity in ideal MHD, then the vertical component of the field must undergo a random walk about zero (if not, then what determines a preferred direction?), such that the BH periodically receives patches of random polarity.  The characteristic timescale between such flips in the mean field would presumably be set by the accretion timescale near the outer radius of the disk $R_{\rm circ}$, which is $t_{\rm visc} \sim 10^3$ s and $\sim 10^6$ s in the case of WD and MS stars, respectively (see \S\ref{sec:flaring}).  Without a large-scale magnetic flux to produce a sustained jet, such polarity flips would cause the jet power to transiently switch off each time the flux changes sign \citep{bhk07,mb09,mtb12} at characteristic intervals $\sim t_{\rm visc}$.  Such large-scale variability is not obviously seen in \ubs\ lightcurve after the first $\sim 10$ days, once it has settled into a magnetically-arrested state.  Instead, the fact that \ubs\ was continuously active over nearly $1.5$ years may suggest that the BH must be threaded by large-scale magnetic flux of the same sign.  Although it is difficult to rule out a dynamo process completely, since the physics of large-scale magnetic field generation in accretion disks is at best poorly understood, we instead focus on the possibility that a reservoir of large-scale magnetic flux is required.

A possible source of large-scale flux is that contained in a pre-existing (`fossil') accretion disk, which was present at the time of TD but was not detectable in pre-imaging of the host of \ubs~since its accretion rate was relatively low (Eddington ratio $\lambda_{\rm fossil}\lesssim10^{-2}$).  During the disruption process, stellar debris is flung outwards onto a series of highly eccentric orbits with an apocenter radius (e.g.~\citealt{Strubbe&Quataert09})
\begin{equation} 
  \frac{r_{\rm
      a}}{r_g}\approx 2\left(\frac{c t}{2\pi
      r_g}\right)^{2/3} \approx 1.3\times10^4 M_{\bh,5}^{-1/3}
  m_\st^{-2/3}r_\st\left(\frac{t}{t_{\rm fb}}\right)^{2/3}. \label{eq:apo} 
\end{equation} 
that increases for material with fall-back times longer than that of the most bound stellar debris $t_{\rm fb}$, where the prefactor in equation (\ref{eq:apo}) is calculated for a solar mass star.  Thus, as debris returns to the BH, it sweeps up a significant fraction\footnote{Although the tidal debris traverses only a small fraction of the azimuthal extent of the disk, the time it spends at pericenter is comparable to the local orbital time.  Thus, a large fraction of the disk will have sufficient time to rotate into, and be collected by, the debris before it falls back.} of the magnetic flux threading the fossil disk at radii $r < r_a(t)$.

We now estimate how luminous the jet from such a fossil disk would
have to be in order to supply the required flux, assuming that the
fossil disk is itself in a MAD state.  In a MAD with a vertical
thickness $h/r\approx0.3{-}0.6$ (as characterizes both super-Eddington
and highly sub-Eddington accretion), every $\Delta r\sim 30\hr r_g$ of the accretion disk contains as much magnetic flux as the BH itself \citep{tm12a,mtb12b}, where $h/r=0.3\hr$.  Thus, the magnetic flux contained by a  magnetically-arrested fossil disk out to a distance $r$ is given by
\citep{tm12a,mtb12}, 
\begin{equation}
  \label{eq:fossilphi}
  \Phi_{\rm D}^{\rm fossil}(r) \approx \frac{r}{30r_g} 
  \left(\frac{\lambda_{\rm fossil}}{\lambda_{\rm MAD}}\right)^{1/2}
  \hr^{-1}
  \Phi_{\bh}^{\rm MAD},
\end{equation}
where we used eq.~\eqref{eq:Phi30} to relate the flux threading the BH by the quiescent disk to that in the MAD state of \ubs, viz.~$\Phi_{\bh}^{\rm fossil} = (\lambda_{\rm fossil}/\lambda_{\rm MAD})^{1/2} \Phi_{\bh}^{\rm MAD}$.
Since the apocenter distance of the infalling debris increases as $r_{a} \propto
t^{2/3}$ (eq.~\ref{eq:apo}), the cumulative amount of ``fallback'' magnetic flux, which is brought to the BH by the infalling tidal streams, is given by
\begin{align}
  \label{eq:fluxoft}
  \Phi^{\rm fb}_{30}(t)&\simeq 0.43 \left(\frac{\lambda_{\rm
        fossil}/\lambda_{\rm MAD}}{10^{-6}}\right)^{1/2}\Phi_{\bh,30}^{\rm MAD}
  M_{\bh,5}^{-1/3} m_\st^{-2/3} r_\star \hr^{-1}\left(\frac{t}{t_{\rm fb}}\right)^{2/3}\\
&\simeq 0.38 \left(\frac{\lambda_{\rm
        fossil}/\lambda_{\rm MAD}}{10^{-6}}\right)^{1/2}
  M_{\bh,5}^{2/3} m_\st^{-2/3} r_* \notag\\
  &\qquad\times\left(\frac{P_{\rm j}}{10^{46}\
      {\rm erg\;s^{-1}}}\right)^{1/2}\hr^{-1}\left(\frac{t}{t_{\rm fb}}\right)^{2/3},
  \label{eq:phimadfb}
\end{align}
where in the last line we have used equation (\ref{eq:Phi30sol}) and have assumed $a = 0.9$.  Figures \ref{fig:fluxmsp} and \ref{fig:fluxwdc} show the time-evolution of $\Phi^{\rm fb}$ in fiducial MS partial disruption and WD scenarios, respectively.  

By demanding in equation (\ref{eq:fluxoft}) that at $t=t_{\rm peak}\simeq1.5t_{\rm fb}$ the accreted flux $\Phi_{\rm fb}$ equal $\Phi_{\bh}^{\rm MAD}$, i.e.~sufficient to magnetically-arrest the disk at $t=t_{\rm MAD}$, this places a lower limit on the Eddington ratio of the fossil disk (using $\lambda_{\rm MAD} \approx 60$):
\begin{align}
  \label{eq:fdlambda}
  \lambda_{\rm fossil}&>
  2\times10^{-4}M_{\bh,5}^{2/3} m_\st^{4/3}r_*^{-2}(t_{\rm peak}/t_{\rm MAD})^{4/3},
\end{align}
corresponding to a jet power of the fossil disk given by (eq.~\ref{eq:Pjet})
\begin{equation}
P_{\rm j}^{\rm fossil}\simmore 3\times10^{40}M_{\bh,5}^{2/3} m_\st^{4/3}r_*^{-2} (t_{\rm peak}/t_{\rm MAD})^{4/3} \,\,{\rm erg\,s^{-1}}.
\end{equation}
If we adopt bolometric and beaming corrections similar to that applied
in~\ubs, then the resulting X-ray luminosity $L_{\rm X,obs}\simmore
5\times 10^{41}M_{\bh,5}^{2/3} m_\st^{4/3}r_*^{-2}$ erg s$^{-1}$ to $t_{\rm MAD} \sim t_{\rm peak}$ is comfortably consistent with {\rm ROSAT} upper limits $L_{\rm X}\simless 2\times 10^{44}$ erg s$^{-1}$ on prior activity from the host galaxy of \ubs~(\citealt{Bloom+11}) for $1\simless M_{\bh,5}\simless 10$.\footnote{That said, it is not at all clear whether the jet from the fossil disk would indeed be pointed along our line of site (= spin axis of the BH) since although the disk is in a MAD state at small radii, the jet direction could be set by the plane of the disk on larger scales, where it is not magnetically-arrested and in general is misaligned with our line of site.}  Thus, a fossil disk sufficiently dim to go undetected prior to \ubs~nevertheless could supply sufficient magnetic flux to power the observed jet.

\subsection{Nature and Duration of the Flaring State}
\label{sec:flaring}

In our model for \ubs, the jets initially point along the rotation axis of the accretion disk, which is misaligned relative to the BH spin and our line of site (Stage 1 in Fig.~\ref{fig:cartoon}).  As $\dot{M}$ decreases, however, magnetic flux accumulated on the BH becomes increasingly dynamically important and the disk enters a magnetically-arrested state, MAD \citep{nia03,tch11}.  Once in a MAD state, strong magnetic fields cause the inner disk and jet to align with the BH spin axis (Stage 2).  However, this alignment process is erratic, with the jets periodically ramming against the accretion flow  \citepalias{mtb12b}.  Occasionally a strong flare is produced when the jet transiently aligns with our line of site, but at most times we observe emission originating from large angles off the core of the jet.  This violent transition to a fully-aligned jet accounts for the extreme variability observed in \ubs~over the first 10 days after the $\gamma$-ray trigger \citep{Bloom+11}.  Since the first evidence for a jet $\sim4$ days prior to the trigger \citep{Krimm+11,Burrows+11}, the total duration of this flaring phase was at least two weeks.

Because the jet alignment is controlled by the magnetic field of the BH,
the characteristic interval between flares is set by the  timescale
for magnetic flux accumulation.  Recent simulations show that large
fluctuations in the jet power ("flares") are produced in MAD flows due
to periodic accumulation and expulsion of flux by the BH on
semi-regular intervals of $(0.5{-}2)10^3M_{\bh,5}$~seconds
\citep{tch11,tm12a}.  To understand this result, note that the inner
$\sim30\hr r_g$ of a MAD accretion disk contain as much flux as the BH
itself \citep{tm12a,mtb12}.  Therefore, when the BH expels an
order-unity fraction of its flux, this flux is replenished on a
characteristic timescale set by the accretion time at $r\simeq15\hr r_g$,
\begin{align}
\Delta t_{\rm flare} \sim t_{\rm acc} &= \alpha_{\rm ss}^{-1}\left(\frac{h}{r}\right)^{-2}\Omega_{\rm K}^{-1}  \nonumber \\ 
&\simeq 3\times 10^{3}{\rm\, s\,}\left(\frac{\alpha_{\rm ss}}{0.1}\right)^{-1}\hr^{-1/2}M_{\bh,5}\left(\frac{r}{15 r_g}\right)^{3/2},
\label{eq:tacc30}
\end{align}
where $\alpha_{\rm ss}$ parameterizes the disk viscosity and $\Omega_{\rm K} = (GM_{\bh}/r^{3})^{1/2}$ is the Keplerian orbital velocity.  Equation (\ref{eq:tacc30}) shows that $\Delta t_{\rm flare}$ is consistent with the observed interval ${\rm  few}\times10^4$~seconds between the large-amplitude flares in \ubs~if the BH is moderately massive, $M_{\bh,5} \gtrsim $ few (depending on $\alpha_{\rm ss}$).

In contrast to the interval between flares, the {\it total} duration of the flaring state must be at least as long as the accretion time scale of the {\it entire}
disk near its outer radius $\sim$ the circularization radius.  Substituting $R_{\rm circ}$ (eq.~[\ref{eq:rd}]) into equation (\ref{eq:tacc30}), we find a flaring duration
\begin{align}
\tau_{\rm acc} &\equiv t_{\rm acc}(r = R_{\rm t}) \nonumber \\ 
&\simeq 5\times 10^{5}{\rm\, s\,}\left(\frac{\alpha_{\rm ss}}{0.1}\right)^{-1}\hr^{-1/2}m_{\star}\beta^{-3/2}\,\,\qquad{\rm [MS]} \\
&\simeq 10^{3}{\rm\, s\,}\left(\frac{\alpha_{\rm ss}}{0.1}\right)^{-1}\hr^{-1/2}(m_{\star}/0.6)^{-1}\beta^{-3/2}.\,\,\,{\rm [WD]}
\label{eq:tauacc}
\end{align}
Again, $\tau_{\rm acc}$ is consistent with the duration of the flaring state in \ubs~($\sim 10^{6}$ s) for a solar-type star with impact parameter $\beta \sim 1$, independent of the BH mass.  However, for a WD $\tau_{\rm acc}$ appears to be too short.

Although $\tau_{\rm acc}$ [eq.~(\ref{eq:tauacc})] sets the minimum duration of the flaring state, the state can last longer if the BH requires more time to fully align the accretion disk with the BH spin.  Alignment completes only once the entire disk is magnetically-arrested.  At the onset of a MAD, most of the [large-scale] magnetic flux in the
system is concentrated near the BH.  As time goes on, two processes
take place.  Firstly, the mass accretion rate drops, causing the
centrally-concentrated magnetic flux to be redistributed to the outer
regions of the accretion disk.  Secondly, new flux is added at the
outer edge of the disk by the infalling stellar debris that has picked
up magnetic flux from the fossil accretion flow (\S\ref{sec:flux}).
Since in a MAD every $\sim30\hr r_g$ in radius holds roughly the same amount magnetic flux as the BH itself \citep{tm12a}, the whole disk goes MAD once its flux reaches a value
\begin{equation}
  \label{eq:diskflux}
  \Phi_{\rm D}^{\rm max} \approx \frac{R_{\rm circ}}{30r_g} \hr^{-1}\Phi_\bh
  \approx 15r_* m_*^{-1/3}M_{\bh,5}^{-2/3}\beta^{-1}\hr^{-1}{\Phi_\bh},
\end{equation}
where $\Phi_{\rm D}$ is the total flux through the midplane of the disk and the last equality makes use of equation \eqref{eq:rd}.  

Now, the flux through the BH evolves as (see eq.~\ref{eq:Phi30}):
\begin{equation}
\Phi_\bh=\Phi_{\bh}^{\rm MAD}(t/t_{\rm MAD})^{-\alpha/2},
\label{eq:phibh}
\end{equation}
where $\Phi_{\bh}^{\rm MAD}$ is BH magnetic flux and $t_{\rm MAD}$ is the time at the
onset of MAD near the hole.  The fallback magnetic flux, brought in by the infalling debris, evolves as (see eq.~\ref{eq:phimadfb}):
\begin{equation}
  \label{eq:phifb}
  \Phi^{\rm fb} = \Phi_\bh^{\rm MAD}\left(\frac{t}{t_{\rm MAD}}\right)^{2/3}.
\end{equation}
 The flux through the disk evolves as
\begin{equation}
  \label{eq:phid}
  \Phi_{\rm D} = (\Phi_{\bh}^{\rm MAD}-\Phi_\bh)+(\Phi^{\rm fb}-\Phi_{\bh}^{\rm MAD}),
\end{equation}
where the disk flux increases as the result of both flux leaving the BH (first term in parentheses) 
and new flux brought in by the infalling debris (second term in parentheses). 

Condition~\eqref{eq:diskflux} can now be written:
\begin{align}
  \label{eq:taualign}
  &\frac{\tau_{\rm align}}{t_{\rm MAD}}  \approx \left(1+\frac{\Phi_{\rm D}^{\rm max}}{\Phi_\bh}\right)^\gamma \approx \left(1+15 r_* m_*^{-1/3}M_{\bh,5}^{-2/3}\beta^{-1}\hr^{-1}\right)^\gamma
  \\
  &\approx  \left[1+0.8 \left(\frac{m_\st}{0.5}\right)^{2/3}\left(\frac{M_{\bh,5}}{3}\right)^{-2/3}\left(\frac{\beta}{2}\right)^{-1}\left(\frac{\hr}{3}\right)^{-1}\right]^{2/3},\;{\rm [MS,\ complete]}\\
  &\approx \left[1+1.9 \left(\frac{m_\st}{0.5}\right)^{2/3}\left(\frac{M_{\bh,5}}{3}\right)^{-2/3}\left(\frac{\beta}{0.8}\right)^{-1}\left(\frac{\hr}{3}\right)^{-1}\right]^{0.57},\;{\rm [MS,\ partial]}\\
  &\approx \left[1+0.4 \left(\frac{m_*}{0.6}\right)^{-2/3}\left(\frac{M_{\bh,5}}{0.1}\right)^{-2/3}\beta^{-1}\left(\frac{\hr}{3}\right)^{-1}\right]^{2/3},\;\;{\rm [WD,\ complete]}
\end{align}
where $\gamma=6/(4+3\alpha)$, and the last three lines are evaluated
for our three progenitor scenarios ($\S\ref{sec:star}$).  Equation
(\ref{eq:taualign}) shows that the jet alignment (flaring) phase can
last for a timescale comparable (or somewhat longer than) than that
required for the MAD to form.  For a MS star, both $\tau_{\rm acc}$
[eq.~\eqref{eq:tauacc}] and $\tau_{\rm align}$ [eq.~\eqref{eq:taualign}]
are sufficiently long to account for the observed duration of the
flaring state in \ubs.  However, for a WD, although $\tau_{\rm acc}$
is very short, the duration of the flaring state is set by $\tau_{\rm
  align} \sim t_{\rm MAD}$ and hence (since $t_{\rm MAD} \sim t_{\rm
  trig} \sim$ days) is also consistent with observations.

Many of the points above are illustrated explicitly in Figures \ref{fig:fluxmsp} and \ref{fig:fluxwdc}, which shows the time-evolution of $\Phi_{\rm D}$ and $\Phi_{\bh}$ in fiducial MS partial disruption and WD scenarios, respectively.  In both cases, $\Phi_{\bh}$ initially rises with the accumulated flux $\Phi_{\rm fb}$, until the inner disk near the BH becomes magnetically arrested.  After this point, flux leaks out of the BH into the surrounding disk and new flux is added by fallback material at the outer edge of the disk ($\Phi_{\rm D}$ rises).  However, eventually $\dot{M}$ drops sufficiently for the entire disk to become magnetically arrested (magnetic field marginally dynamically-important everywhere).  At this point, even the disk itself cannot hold the accumulated flux, which begins to leak out, causing $\Phi_{\rm D}$ to fall.  The jet alignment/flaring phase described above (Stage 2) occurs during the interval when $\Phi_{\rm D}$ is still rising.

\subsection{Origin of Radio Rebrightening}
\label{sec:radio}

Our fits to the X-ray light curve of \ubs~(Figs.~\ref{fig:Fx}, \ref{fig:tdc}, \ref{fig:tdp}) show that the jet could have been active weeks-month prior to the first detection.  However, since the jet was pointed away from$-$and possibly precessing about$-$our line of site (= spin axis of the BH), its emission was not initially observable due to geometric or relativistic beaming.  Nevertheless, since the jet still injects relativistic kinetic energy into the surrounding ISM during this phase, this gives rise to delayed radio afterglow emission on a timescale of months$-$year for a typical misalignment angle ($\S\ref{sec:radio_rebrightening}$), consistent with observed radio re-brightening several months after the trigger (B12).

To produce radio rebrightening of the magnitude observed in \ubs, the additional energy released by the early off-axis jet must
exceed that released later during the on-axis phase (i.e.~after the initial $\gamma$-ray detection).  Since the mass accretion rate
decreases as $\dot{M} \propto t^{-\alpha}$ and the jet power $P_{\rm
  j} \propto \dot{M}$ [eq.~(\ref{eq:Pjet})] during MAD accretion, then
the maximum\footnote{This is a maximum since we assume that
  accretion is in a MAD state at all times, such that $P_{j} \sim
  \dot{M}$c$^{2}$ (eq.~[$\ref{eq:Pjet}]).$  Prior to the onset of MAD
  (Stage 1 in Fig.~\ref{fig:cartoon}) the jet efficiency is instead
  limited by the amount of accumulated magnetic flux
  (eq.~[\ref{eq:pj}]).  Also, the jet can be stifled at early times
  when the magnetic field is dynamically weak ($\phi_\bh<\phi_{\rm
    on}\approx20$) by the high ram pressure of the thick misaligned inflow (\citealt{kb09}).} energy released by the jet prior to time $t$
is given by 
\begin{equation}
  \label{eq:pretriggerpower}
  E(t)\propto \int_{t_{\rm peak}}^{t}t'^{-\alpha}dt' = \frac{t_{\rm peak}^{1-\alpha}-t^{1-\alpha}}{\alpha-1},
\end{equation}
where $\alpha\sim5/3{-}2.2$. If one demands that the jet energy released
before the trigger exceeds the energy released after the trigger,
\begin{equation}
  \label{eq:encond}
  1<\frac{E(t_{\rm trig})}{E(\infty)-E(t_{\rm trig})} = \left(\frac{t_{\rm
      trig}}{t_{\rm peak}}\right)^{\alpha-1}-1,
\end{equation}
then this requires
\begin{equation}
  \label{eq:tcond}
  t_{\rm trig}> 2^{1/(\alpha-1)} t_{\rm peak}.
\end{equation}
Aside from a different prefactor, this constraint is identical to that
based on the shape of the X-ray light curve
(eq.~[\ref{eq:tpeakcond}]).  Self-consistency of our model thus
requires that the off-axis power be similar to that required to
explain the observed radio rebrightening.  

Although we focus above on energy injected off-axis prior to the $\gamma$-ray trigger, this is not the only means by which the jets could inject ``invisible'' energy into the ambient medium.  The onset of MAD accretion after the $\gamma$-ray trigger may cause the jet to wobble in and out of our line of sight, giving rise to high amplitude variability ($\S\ref{sec:wobbling_jet}$; Stage 2 in Fig.~\ref{fig:cartoon}).  Since during this process the jets are pointed towards our line of site only a fraction of the time, most of their energy is released during a mis-aligned state.  Indeed, the $\sim 10$ per cent duty cycle of the observed flaring (Fig.~\ref{fig:Fx}) suggests that the energy injected during misaligned phases could exceed that injected along our line of site by an order of magnitude.  Since $\sim 1/2$ of the total X-ray fluence occurred during the first $t-t_{\rm trig}\simless 10$ days (wobbling jet phase), misaligned jets could produce off axis relativistic ejecta with $\sim 5$ times more energy than that directly probed by the observed $\gamma$-ray/X-ray emission.  This alone might be sufficient to power the observed rebrightening, without the need for significant energy injection prior to the trigger.

\subsection{Predictions}
\label{sec:predictions}
\subsubsection{X-ray Transients}

We postulate that \ubs~resulted from a somewhat special geometric
configuration, in which the BH spin axis was pointed along our line of
site.  This favourable geometry resulted in several bright flares over
the first $\sim$ 10 days, which are produced as the jet settles down
into its fully aligned configuration (Stage 2 in
Fig.~\ref{fig:cartoon}; \S~\ref{sec:trigger}, \ref{sec:flaring}).  If,
however, we had instead been positioned along a more `typical' line of
site not aligned with the BH, then we might only have observed a small
number of flares, possibly just one in the majority of cases.  Such a
single flare would appear as an X-ray/soft $\gamma$-ray transient of
duration $\sim10^3$~s, yet unaccompanied by an extended luminous X-ray
tail as characterized \ubs.

Could such a population of off-axis jetted TDE flares be contributing
to the known population of high energy transients?  Given their long
duration and low luminosity, such flares could be mistaken for a
long-soft gamma-ray burst (GRB) or an X-ray Flash.  Perhaps the
closest known analog is the class of `low-luminosity GRBs' (LLGRBs)
(e.g.~\citealt{Soderberg+06}; \citealt{Cobb+06}), although most of
these are probably not standard TD events due to their locations off
the nucleus of their host galaxies and their observed association with
core-collapse supernovae (e.g.~\citealt{Chornock+10}; but see
\citealt{Shcherbakov+12}).  The volumetric rate of LLGRBs actually
exceeds that of classical GRBs (\citealt{Soderberg+06}), but only a
handful are known since they are more challenging to detect than
luminous high redshift GRBs (see \citealt{Nakar&Sari12} for a recent
compilation).  Thus, even if the rate of single-flare off-axis TD jets
is a factor of $\sim 10$ times higher than the rate of \ubs-like
events, it is unclear how many should have been detected yet.  If such
an event is eventually detected, perhaps by the next generation of
wide-field X-ray telescopes, it may be distinguished from normal
LLGRBs by its [1] nuclear position; [2] associated optical/UV/soft
X-ray emission produced by isotropic thermal emission from the
accretion disk or ionized stellar ejecta
(\citealt{Strubbe&Quataert11},\citealt{Clausen+12}), rather than a
core-collapse SNe; [3] delayed radio emission from the off-axis jet
(\citealt{Giannios&Metzger11}; \S\ref{sec:radiotransients}).

\subsubsection{Jet Revival}
Although the jet in \ubs~is presently off, our model predicts that the
BH continues to accrete through a geometrically-thin accretion disk in
a jet-less ``thermal state''.  However, eventually the mass accretion
rate will decrease below $\approx 2\%$ of $\dot M_{\rm Edd}$
\citep{2003A&A...409..697M}, after which point the flow may transition
to a ``hard'' state, as characterized by a radiatively-inefficient
geometrically-thick accretion flow.  Once this transition occurs,
magnetic flux can once again accumulate near the black hole on a short
timescale ($\sim t_{\rm acc}$; eq.~\ref{eq:tauacc}), resulting in a
new MAD accretion phase (Stage 5 in Fig.~\ref{fig:cartoon}).  A jet
aligned with the Earth and BH spin axis will thus again form, with a
power that again tracks the accretion rate $\dot M$.  From Figures
\ref{fig:tdc} and \ref{fig:tdp} we estimate that the X-ray flux will
be $\sim 2\times10^{-14}-10^{-13}$ erg cm$^{-2}$ s$^{-1}$ when the jet
turns back on, well within the detection limits of current X-ray
observatories.

In order to test this idea, we strongly encourage regular X-ray
follow-up of \ubs~over the next decade.  The timescale and flux of the
observed revival would inform [1] the rate at which the accretion rate
is decreasing, and therefore help to distinguish between partial and
complete disruption scenarios, or whether the disk has transitioned to a
spreading evolution (e.g.~\citealt{Cannizzo+11}) [2] the ratio of
accretion rates which characterize the thick$\rightarrow$thin disk and
thin$\rightarrow$thick disk transitions, respectively; [3] and whether
the jet beaming correction (related to the opening angle and bulk
Lorentz factor) during the low-hard state are similar to those during
the super-Eddington state.

\subsubsection{Ubiquity of Radio Transients from TD Events}
\label{sec:radiotransients}

Due to the enormous energy released in relativistic ejecta, \ubs~remains a bright radio source ($F_\nu \sim 1-10$ mJy at $\nu \sim 1-50$ GHz) even now, almost two years after the TD event.  Since most of the jet activity occurred $\sim$ months around the trigger time, by now the ejecta has slowed down appreciably due to interaction with the circumnuclear medium.  The current expansion velocity of the ejecta is at most mildly relativistic $\gamma \sim 2$ (B12), in which case the radio emission should be relatively isotropic (i.e.~flux varying by less than an order of magnitude from front to side).  

The fact that \ubs~is a bright isotropic radio source has two implications: First, given the relatively close proximity ($z \sim 0.1$) of most candidate events detected over the past several years, any jet from these events as remotely as powerful as that in \ubs~should be easily detectable by now, even if the jet remains always pointed away from our line of site.  \citet{Bower+12} and \citet{VanVelzen+12} have conducted radio follow-up observations of previous thermal TD flares; since most of these observations produced only deep upper limits (with a few interesting exceptions), this already constrains the fraction of TD events accompanied by powerful jets to be $\lesssim 10$ per cent.  In hindsight it is perhaps unsurprising that \ubs~is unique, given the enormous magnetic flux required to power its jet, which could require special conditions not satisfied by most TDs ($\S\ref{sec:flux}$; \citealt{DeColle+12}).  A second consequence of the radio luminosity of \ubs~is that off-axis emission from other jetted TD events (albeit rare) is isotropic and hence should be detectable out to much higher redshifts ($z \sim 1$).  Such events are one of the most promising sources for future wide-field radio surveys (\citealt{Giannios&Metzger11}; \citealt{VanVelzen+11}; \citealt{Frail+12}; see \citealt{Cenko+12} for a potential high redshift analog to \ubs), which will help constrain the rate and diversity of jetted TD events.

\section{Conclusions}
\label{sec:conclusions}

We have presented a self-consistent model that explains many of the
previously disparate puzzles associated with the jetted TD event
\ubs~(Fig.~\ref{fig:cartoon}; $\S\ref{sec:scenario}$).  This model
relies on just one major assumption: the accumulation of a large,
dynamically-important magnetic flux near the central BH, such that the
accretion flow from the returning stellar debris becomes
magnetically-arrested (MAD; \S\ref{sec:wobbling_jet}) on a timescale
of $\sim$ week-month after the TD event
(Figs.~\ref{fig:tdc}--\ref{fig:tdp}).  The onset of MAD accretion in
\ubs~naturally accounts for (i) the period of intense flaring that
lasted for the first $\sim10$ days after the trigger; (ii) the
approximate constancy (in a time-average sense) of the luminosity
during this period; (iii) the subsequent transition of X-ray
luminosity to a steady (non-precessing) jet with a power that tracks
the predicted power-law decline in the accretion rate; (iv) the sudden
shut off of the jet emission at $\sim500$ days after the trigger; and
(v) the potential origin of the mysterious late-time radio
rebrightening that started about a month after the trigger.  Our model
also naturally predicts a QPO (at a frequency closely tied to that of
a BH spin) that is consistent with that seen in the lightcurve of
\ubs.

We emphasize that the strong magnetic flux required by our model is
not an entirely independent assumption: a flux of at least this
magnitude is necessary to explain the observed jet power in the first
place (eq.~\ref{eq:pj}; $\S\ref{sec:flux}$; given plausible values of
the BH mass and spin), while a much weaker flux would be unable to
produce a jet at all (the jet would be `stifled') against the powerful
ram pressure of the misaligned disk \citep{kb09}.

\acknowledgements

We thank James Guillochon, Michael Kesden, Serguei Komissarov, Julian Krolik, Morgan
MacLeod, Jonathan C.~McKinney, Petar Mimica, Ramesh Narayan, Ryan
O'Leary, Asaf Pe'er, Tsvi Piran, Eliot Quataert, Roman Shcherbakov, 
Nicholas Stone, and Sjoert van Velzen for insightful discussions. AT
was supported by a Princeton Center for Theoretical Science fellowship
and an XSEDE allocation TG-AST100040 on NICS Kraken and Nautilus and
TACC Ranch.



{\small
\bibliographystyle{mn2e}

}
\label{lastpage}
\end{document}